%% file: main.tex
\documentclass[
reprint, % use for submission
superscriptaddress,
 amsmath,
 amssymb,
 aps,
prb,
]{revtex4-2}

\usepackage{placeins}
\usepackage[percent]{overpic}
\usepackage{cuted}

\usepackage{graphicx}
\graphicspath{ {./figures/} }

\usepackage{dcolumn} % Align table columns on decimal point
\usepackage{bm}% bold math
\usepackage[colorlinks=true,
            linkcolor=blue,
            citecolor=blue,
            urlcolor=magenta]{hyperref}

\usepackage[mathlines]{lineno} % Enable numbering of text and display math
\begin{document}

\preprint{APS/123-QED} % ???

\title{Electronic Reconstruction across the Tilt-Free Transition in La$_3$Ni$_2$O$_7$}

\author{Mengjie Kong}
\affiliation{SOLEIL Synchrotron, L'Orme des Merisiers, RD 128, Saint Aubin 91190, France}
\affiliation{Université Paris-Saclay, 3 rue Joliot Curie, Bâtiment Breguet, 91190 Gif-sur-Yvette, France}

\author{Gergely N\'emeth}
\affiliation{SOLEIL Synchrotron, L'Orme des Merisiers, RD 128, Saint Aubin 91190, France}

\author{Yingpeng Yu}
\affiliation{Beijing National Laboratory for Condensed Matter Physics, Institute of Physics, Chinese Academy of Sciences, Beijing 100190, China}
\affiliation{School of Physical Sciences, University of Chinese Academy of Sciences, Beijing 100190, China}

\author{Bosen Wang}
\affiliation{Beijing National Laboratory for Condensed Matter Physics, Institute of Physics, Chinese Academy of Sciences, Beijing 100190, China}
\affiliation{School of Physical Sciences, University of Chinese Academy of Sciences, Beijing 100190, China}

\author{Jianping Sun}
\affiliation{Beijing National Laboratory for Condensed Matter Physics, Institute of Physics, Chinese Academy of Sciences, Beijing 100190, China}
\affiliation{School of Physical Sciences, University of Chinese Academy of Sciences, Beijing 100190, China}

\author{Jinguang Cheng}
\affiliation{Beijing National Laboratory for Condensed Matter Physics, Institute of Physics, Chinese Academy of Sciences, Beijing 100190, China}
\affiliation{School of Physical Sciences, University of Chinese Academy of Sciences, Beijing 100190, China}

\author{Ferenc Borondics}
\affiliation{SOLEIL Synchrotron, L'Orme des Merisiers, RD 128, Saint Aubin 91190, France}

\author{Bastien Michon}
\thanks{Corresponding author: \href{mailto:bastien.michon@univ-tours.fr}{bastien.michon@univ-tours.fr}}
\affiliation{SOLEIL Synchrotron, L'Orme des Merisiers, RD 128, Saint Aubin 91190, France}
\affiliation{Present address: GREMAN - UMR7347 CNRS, Université de Tours, INSA Centre Val de Loire, Parc de Grandmont, Tours 37200, France}

%\date{\today}% It is always \today, today, but any date may be explicitly specified

\begin{abstract}
The emergence of high-$T_c$ superconductivity in pressurized La$_3$Ni$_2$O$_7$ is intimately linked to a structural transition that suppresses the tilts of the NiO$_6$ octahedra, yet its impact on the electronic structure remains poorly understood. Here, we probe the electronic response across this tilt-free transition at $T_{\mathrm{st}}\simeq544$~K using broadband infrared-to-visible reflectivity at ambient pressure, covering photon energies from 15~meV to 3.2~eV. We observe a pronounced redistribution of spectral weight over an exceptionally broad energy range, with spectral weight transferred from excitations between 1 and 3~eV toward low-energy excitations below 1~eV. Most strikingly, two low-energy interband excitations progressively converge and merge upon entering the tilt-free phase, revealing a substantial reconstruction of the finite-energy electronic structure. These changes point to a reconstruction of the bilayer Ni $3d_{z^2}$-derived electronic states, whose interlayer coupling is central to proposed mechanisms of superconductivity in La$_3$Ni$_2$O$_7$. Our results establish the tilt-free transition as a direct route to reorganizing the electronic degrees of freedom implicated in high-$T_c$ superconductivity and provide an ambient-pressure reference for the electronic structure of the superconducting state.

% \begin{description}
% Secondary publications and information retrieval purposes.
% You may use the \texttt{description} environment to structure your abstract;
% use the optional argument of the \verb+\item+ command to give the category of each item. 
% \end{description}
\end{abstract}

\maketitle

%%% this is where things happen

\input{manuscript}

\FloatBarrier
\bibliography{biblio}

\end{document}

\preprint{APS/123-QED} % ???

\title{Supplementary: Electronic Reconstruction across the Tilt-Free Transition in La$_3$Ni$_2$O$_7$}

\author{Mengjie Kong}
\affiliation{SOLEIL Synchrotron, L'Orme des Merisiers, RD 128, Saint Aubin 91190, France}
\affiliation{Université Paris-Saclay, 3 rue Joliot Curie, Bâtiment Breguet, 91190 Gif-sur-Yvette, France}

\author{Gergely N\'emeth}
\affiliation{SOLEIL Synchrotron, L'Orme des Merisiers, RD 128, Saint Aubin 91190, France}

\author{Yingpeng Yu}
\affiliation{Beijing National Laboratory for Condensed Matter Physics, Institute of Physics, Chinese Academy of Sciences, Beijing 100190, China}
\affiliation{School of Physical Sciences, University of Chinese Academy of Sciences, Beijing 100190, China}

\author{Bosen Wang}
\affiliation{Beijing National Laboratory for Condensed Matter Physics, Institute of Physics, Chinese Academy of Sciences, Beijing 100190, China}
\affiliation{School of Physical Sciences, University of Chinese Academy of Sciences, Beijing 100190, China}

\author{Jianping Sun}
\affiliation{Beijing National Laboratory for Condensed Matter Physics, Institute of Physics, Chinese Academy of Sciences, Beijing 100190, China}
\affiliation{School of Physical Sciences, University of Chinese Academy of Sciences, Beijing 100190, China}

\author{Jinguang Cheng}
\affiliation{Beijing National Laboratory for Condensed Matter Physics, Institute of Physics, Chinese Academy of Sciences, Beijing 100190, China}
\affiliation{School of Physical Sciences, University of Chinese Academy of Sciences, Beijing 100190, China}

\author{Ferenc Borondics}
\affiliation{SOLEIL Synchrotron, L'Orme des Merisiers, RD 128, Saint Aubin 91190, France}

\author{Bastien Michon}
\thanks{Corresponding author: \href{mailto:bastien.michon@univ-tours.fr}{bastien.michon@univ-tours.fr}}
\affiliation{SOLEIL Synchrotron, L'Orme des Merisiers, RD 128, Saint Aubin 91190, France}
\affiliation{Present address: GREMAN - UMR7347 CNRS, Université de Tours, INSA Centre Val de Loire, Parc de Grandmont, Tours 37200, France}

\maketitle

%%% this is where things happen

\input{Supplementary}

%% file: manuscript.tex
%\tableofcontents

\textbf{\textit{Introduction}}\ -- The recent discovery of high-$T_c$ superconductivity near 80~K in pressurized La$_3$Ni$_2$O$_7$ has attracted considerable attention~\cite{Sun23,GWang24,YZhang24,JLi25,FLi26}. Particular interest has been focused on the concomitant structural evolution, characterized by the suppression of the tilts in NiO$_6$ oxygen octahedra transforming the low-symmetry tilted \textit{Amam} phase toward a higher-symmetry tilt-free structure (\textit{Fmmm} or \textit{I4/mmm})~\cite{LWang24,Geisler24,HZhang25}. In the T-P phase diagram, the \textit{Amam} phase is bounded along both the pressure and temperature directions, at around 15~GPa at room temperature and $T_{\mathrm{st}}\simeq544$~K at ambient pressure~\cite{BMichon26}.

%However, recent experiments indicate that the realization of a particular high-symmetry structure is not a prerequisite for superconductivity in La$_3$Ni$_2$O$_{7-\delta}$~\cite{MShi25}. This suggests that crystallographic symmetry alone cannot account for the emergence of superconductivity, raising the question of how the electronic structure evolves in response to pressure and the accompanying structural changes. \textbf{Je ne comprends pas ce paragraphe et la logique derrière ? Pour moi c'est hors sujet.}

The electronic structure of La$_3$Ni$_2$O$_7$ is strongly influenced by its characteristic bilayer geometry. The two NiO$_2$ layers within each bilayer are connected through inner apical oxygen atoms, which mediate interlayer coupling between Ni $3d_{z^2}$ orbitals and give rise to bonding and antibonding states~\cite{BGeisler24}. The relevance of these states to the low-energy electronic structure has also been demonstrated by angle-resolved photoemission spectroscopy (ARPES)~\cite{JY24}. Consequently, the electronic reconstruction across the tilt-free transition may depend not only on the change in crystallographic symmetry, but also on the evolution of local structural parameters that modify orbital overlap and interlayer coupling~\cite{Geisler24}.

Theoretical calculations predict substantial changes in both the Drude (free-carrier) and interband optical responses across the tilt-free transition. Experimentally, high-pressure reflectivity reveals a pronounced enhancement of the Drude spectral weight, while a reversible color change of the sample points to a concomitant reconstruction of the interband response in the visible range~\cite{BGeisler24,BMichon26}. However, because of the high refractive index and strong absorptions of the diamond in the mid-infrared region, the interface between the diamond-anvil-cell and the sample limits a quantitative assessment of how the interband responses are modified through the structural transition. The possibility to drive the tilt-free transition at ambient pressure upon heating up to $T_{\mathrm{st}}\simeq544$~K \cite{Koba96,Sasa97,Zink07,BMichon26} therefore offers a complementary route to probe the electronic reconstruction inside the tilt-free phase.

Here, we investigate the temperature-driven electronic reconstruction in La$_3$Ni$_2$O$_7$ through high-temperature (HT) reflectivity microspectroscopy measurements covering a broad photon-energy range from 15~meV (far-infrared with synchrotron source) to 3.2~eV (visible). By tracking the reflectivity across the structural transition at $T_{\mathrm{st}}\simeq544$~K, we reveal a pronounced redistribution of spectral weight over a broad energy range, together with substantial changes in the low-energy electronic response associated to interband excitations.

\begin{figure}[h!]
\includegraphics[scale=0.55]{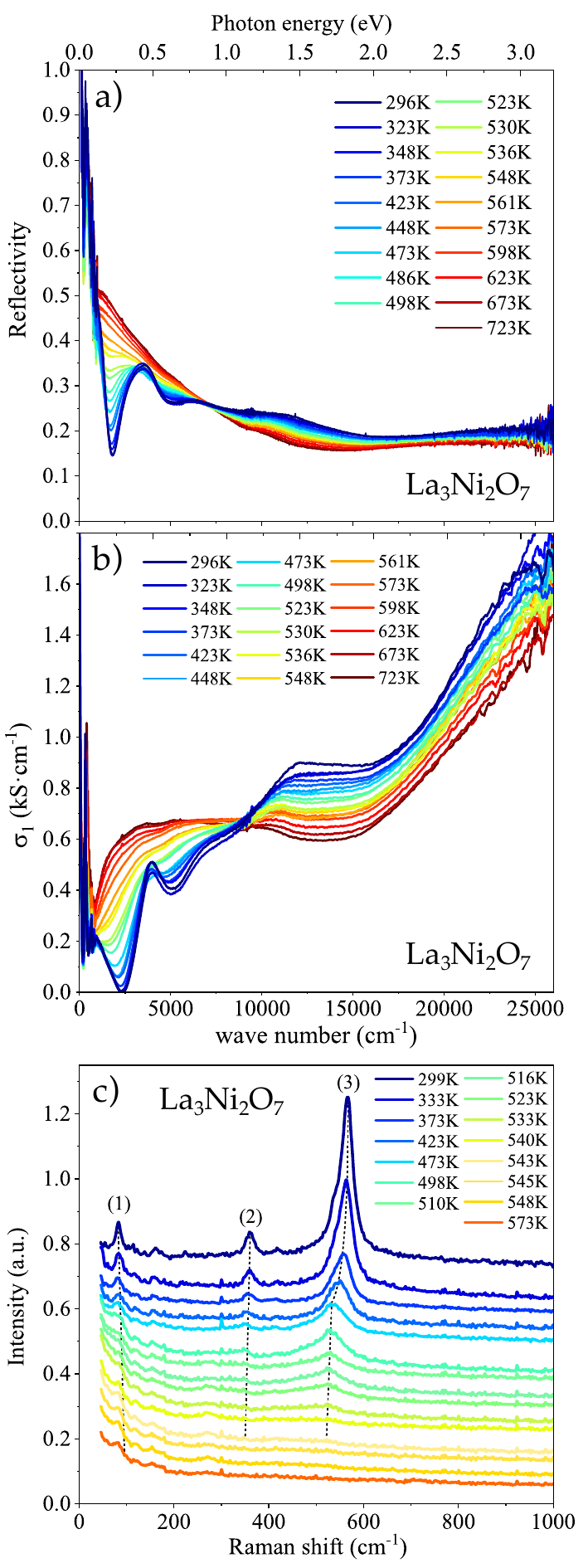}
\caption{\label{fig:fig1} High-temperature infrared and Raman spectroscopies data across the structural transition at $T_{\mathrm{st}}\simeq544$~K. (a) HT Reflectivity data as a function of photon frequency/energy $R(\omega)$ obtained from different set of measurements with energies varying from 15~meV to 3.2~eV. (b) Real part of the optical conductivity $\sigma_1(\omega)$ obtained from Kramers-Kronig analysis of $R(\omega)$ data in (a). (c) Complementary HT Raman data underlining the concomitant structural transition around $T_{\mathrm{st}}\simeq544$~K with the suppression of modes labeled (2) and (3) corresponding to tilt-shearing modes in NiO$_6$ octahedra \cite{HZhang25}.}
\end{figure}

\par\medskip
\noindent\textbf{\textit{Experimental data}}\ -- We performed HT reflectivity measurements at ambient pressure on the \textit{ab} plane of La$_3$Ni$_2$O$_7$ mounted on a heating stage to precisely control the temperature from 296 to 723~K. Broadband spectra were constructed by combining measurements over four complementary spectral ranges (see Methods~\cite{suppmat}). The reflectivity spectra were calibrated thanks to gold (infrared) and aluminum (near-infrared and visible) references measured at each temperature. The measurements were reproduced over several heating and cooling cycles, confirming the reproducibility of the observed spectral evolution and the absence of sample degradation. Fig.~\ref{fig:fig1}(a) shows the resulting reflectivity $R(\omega)$ at different temperatures. $R(\omega)$ approaches unity towards the low-frequency limit, consistent with the metallic nature of the material. At low temperatures (blue curves), a broad depletion is observed around 2000~cm$^{-1}$. This feature progressively weakens upon heating and is strongly suppressed around the structural transition at $T_{\mathrm{st}}\simeq544$~K (yellow and orange curves), where $R(\omega)$ becomes smooth. The spectra cross around 8000~cm$^{-1}$ ($\sim$~1~eV), above which $R(\omega)$ decreases with increasing temperature over a broad spectral range extending to approximately 26,000~cm$^{-1}$ ($\sim$~3.2~eV). The largest variation occurs around 12,500~cm$^{-1}$ ($\sim$~800~nm), corresponding to the red part of the visible spectrum and accounting for the previously reported color change of the samples~\cite{BMichon26}. Overall, the distinct temperature evolution of the low- and high-energy responses reveals a pronounced reshaping of the optical response over a broad energy range.

\begin{figure*}[htp!]
\includegraphics[scale=0.45]{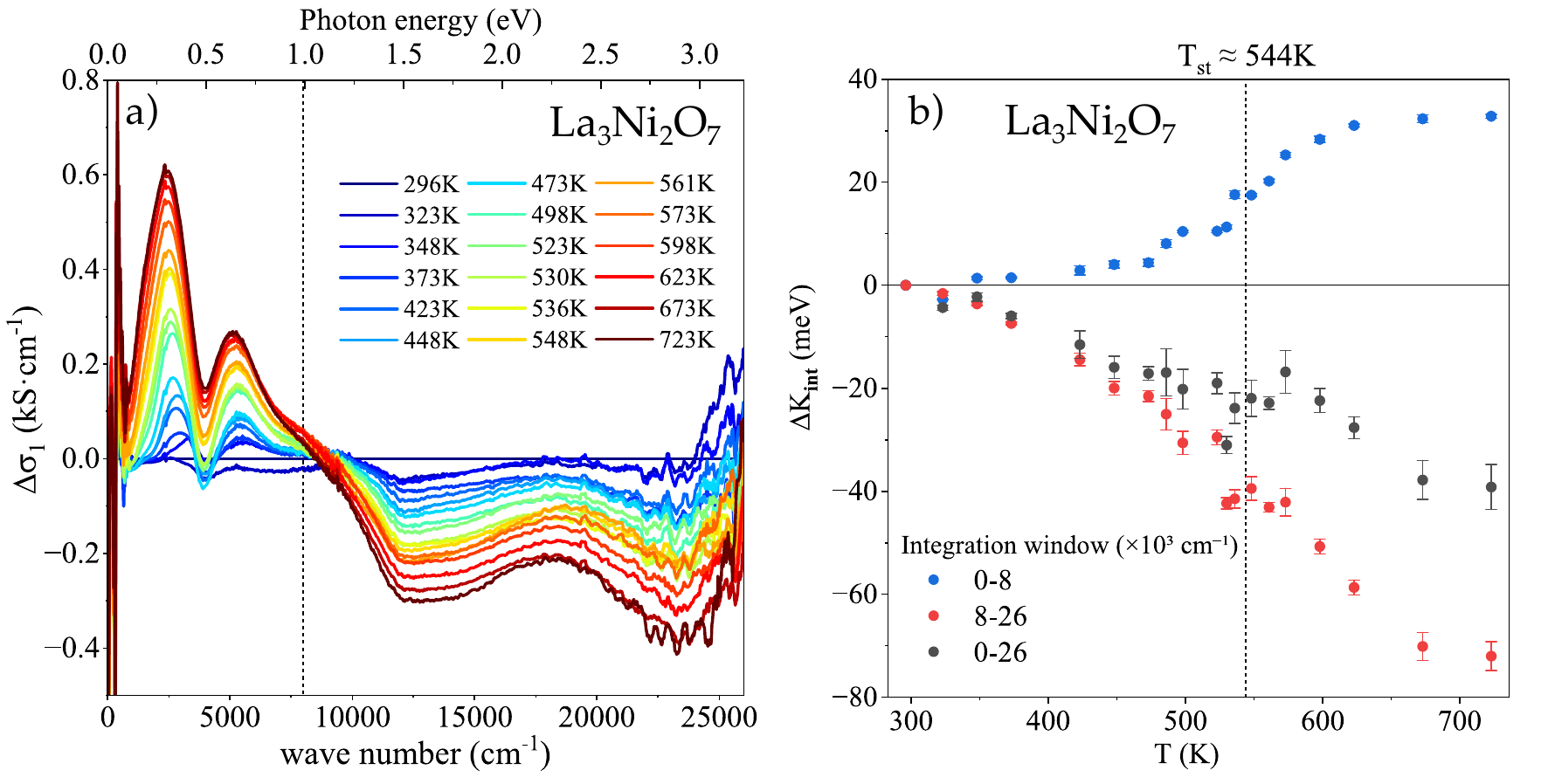}
\caption{\label{fig:fig2} Spectral weight redistribution evidences from HT spectroscopic data. (a) The differential optical conductivity data $\Delta\sigma_1(\omega,T)$ relative to 296~K underlines a pronounced spectral weight redistribution at high temperature from negative region (high frequency range) to positive (low frequency). (b) Relative spectral weight $\Delta K_{\mathrm{int}}(T)$ as a function of temperature obtained from the integration of $\Delta\sigma_1(\omega,T)$ on different spectral ranges: 0--8000 (in blue circles), 8000--26\,000 (red circles), and 0--26\,000~cm$^{-1}$ (dark gray). Across $T_{\mathrm{st}}\simeq544$~K, the low frequency range 0--8000~cm$^{-1}$ remarkably gains spectral weight.}
\end{figure*}

The corresponding real part of the optical conductivity, $\sigma_1(\omega)$, obtained from Kramers-Kronig analysis of the reflectivity spectra $R(\omega)$ (see Methods~\cite{suppmat}), is shown in Fig.~\ref{fig:fig1}(b) (in Fig.~S1(a) for the imaginary part $\sigma_2(\omega)$). At low temperatures (blue curves), several distinct structures are resolved in the mid-infrared region, including features around 1000 and 4000~cm$^{-1}$. Upon heating, $\sigma_1(\omega)$ increases markedly across this range, progressively filling the spectral-weight depletion observed at low temperatures. At higher energies, the conductivity exhibits a broad shoulder around 12\,000~cm$^{-1}$, followed by a pronounced band centered near 25\,000~cm$^{-1}$. Both features progressively decrease with increasing temperature. The opposite temperature evolution of the low- and high-energy responses suggests a substantial redistribution of the electronic spectral weight over a broad energy range.

Complementary HT Raman measurements were performed at ambient pressure using a 532~nm laser, with the sample mounted on the same heating stage (see Methods~\cite{suppmat}). Fig.~\ref{fig:fig1}(c) shows the resulting Raman spectra between 299 and 573~K. Three characteristic modes are observed around 90, 360, and 565~cm$^{-1}$, labeled (1), (2), and (3), respectively, and tracked by black dotted lines. These Raman modes progressively redshift upon heating, consistent with thermal lattice expansion. Upon heating, mode (1) progressively develops an asymmetric Fano-like line shape, while modes (2) and (3), associated with tilts of the NiO$_6$ octahedra \cite{HZhang25}, exhibit a similar asymmetry before disappearing completely between 543 and 545~K (yellowish curves in Fig.~\ref{fig:fig1}(c)), coincident with the structural transition at $T_{\mathrm{st}}\simeq544$~K, as previously reported~\cite{BMichon26}. The increasing Fano-like asymmetry reflects an enhanced interplay between the phonon excitations and the electronic continuum~\cite{UFano61,TDeve07}. A similar evolution of the Fano response has been reported across the temperature- and pressure-driven structural transitions in La$_3$Ni$_2$O$_7$, where it accompanies an enhanced Drude response (free electrons) and a crossover toward a more metallic state~\cite{BMichon26}. 

The concomitant evolution observed in Figs.~\ref{fig:fig1}(a--c) therefore indicates that the modification of electronic responses are closely coupled to changes in the lattice structure.

\par\medskip
\noindent\textbf{\textit{Spectral weight redistribution}}\ -- To further analyze the temperature-induced evolution of the optical response, we calculate the differential optical conductivity relative to 296~K (room temperature),
\begin{equation}
\Delta\sigma_1(\omega,T) = \sigma_1(\omega,T)-\sigma_1(\omega,296~\mathrm{K}),
\end{equation}
as shown in Fig.~\ref{fig:fig2}(a). Below 8000~cm$^{-1}$ ($\sim$~1~eV), the positive contributions increase upon heating, while the minima around 800 and 3900~cm$^{-1}$ evolve from negative to positive values near $T_{\mathrm{st}}\simeq544$~K (yellow and orange curves). In contrast, above 8000~cm$^{-1}$, $\Delta\sigma_1(\omega)$ becomes progressively more negative with increasing temperature. The sign change of the minima around 800 and 3900~cm$^{-1}$ indicates a pronounced reshaping of the low-energy optical responses across $T_{\mathrm{st}}$, rather than a uniform increase in conductivity. Since such structures in $\Delta\sigma_1(\omega)$ can arise from changes in the spectral weight, resonance frequency, or linewidth of the underlying excitations, the evolution of the individual optical components is examined afterwards using a Drude--Lorentz decomposition.

To quantify this spectral-weight redistribution, we integrate $\Delta\sigma_1(\omega,T)$ over a selected frequency range $[\omega_1,\omega_2]$,
\begin{equation}
\Delta K_{\mathrm{int}}([\omega_1,\omega_2],T) = \frac{2d_c\hbar^2}{\pi e^2} \int_{\omega_1}^{\omega_2} \Delta\sigma_1(\omega,T)\,d\omega ,
\end{equation}
where $\Delta K_{\mathrm{int}}(T) = K_{\mathrm{int}}(T) - K_{\mathrm{int}}(296~K)$ represents the relative spectral weight variation and $d_c = 5.1$~\AA \ the mean interlayer spacing in La$_3$Ni$_2$O$_7$~\cite{BMichon21}. In Fig.~\ref{fig:fig2}(a), the boundary at 8000~cm$^{-1}$ ($\sim$~1~eV) in vertical dotted line approximately separates the low-energy spectral-weight gain from the predominantly negative high-energy response in $\Delta\sigma_1(\omega)$. It also corresponds to the crossing region observed in the reflectivity data (Fig.~\ref{fig:fig1}(a)). 

Fig.~\ref{fig:fig2}(b) shows $\Delta K_{\mathrm{int}} (T)$ integrated over 0--8000 (in blue circles), 8000--26\,000 (red circles), and 0--26\,000~cm$^{-1}$ (dark gray). The spectral weight below 8000~cm$^{-1}$ increases with temperature and exhibits a pronounced step-like enhancement around $T_{\mathrm{st}}$. In contrast, the spectral weight between 8000 and 26,000~cm$^{-1}$ continuously decreases upon heating without any marked change across the structural transition, and is not compensated by the low-energy variation (gray points for $K_{\mathrm{int}} (T)$ between 0 and 26,000~cm$^{-1}$). Because this loss is only partially recovered at lower energies, the redistribution must extend to higher energies in the ultraviolet. The structural transition therefore drives a broad reorganization of the electronic spectral weight over an energy scale of several electronvolts.

The cumulative spectral-weight change $\Delta K_{\mathrm{int}}(\omega)$ defined with the interval $[\omega_1,\omega_2] = [0,\omega]$ (see Fig.~S2b and Methods~\cite{suppmat}) reaches a maximum around 9000--10\,000~cm$^{-1}$ before decreasing and remaining negative at 26\,000~cm$^{-1}$, consistent with the partial spectral weight redistribution. A decomposition into 1000--2000~cm$^{-1}$ frequency intervals (Fig.~S3) further shows that the redistribution involves multiple energy ranges, with spectral-weight gains at lower energies and losses at higher energies, and becomes particularly pronounced across $T_{\mathrm{st}}$. This finer decomposition also shows that the overall redistribution is robust against the specific choice of 8000~cm$^{-1}$ as the integration boundary.

\par
\begin{figure}[h!]
\includegraphics[scale=0.42]{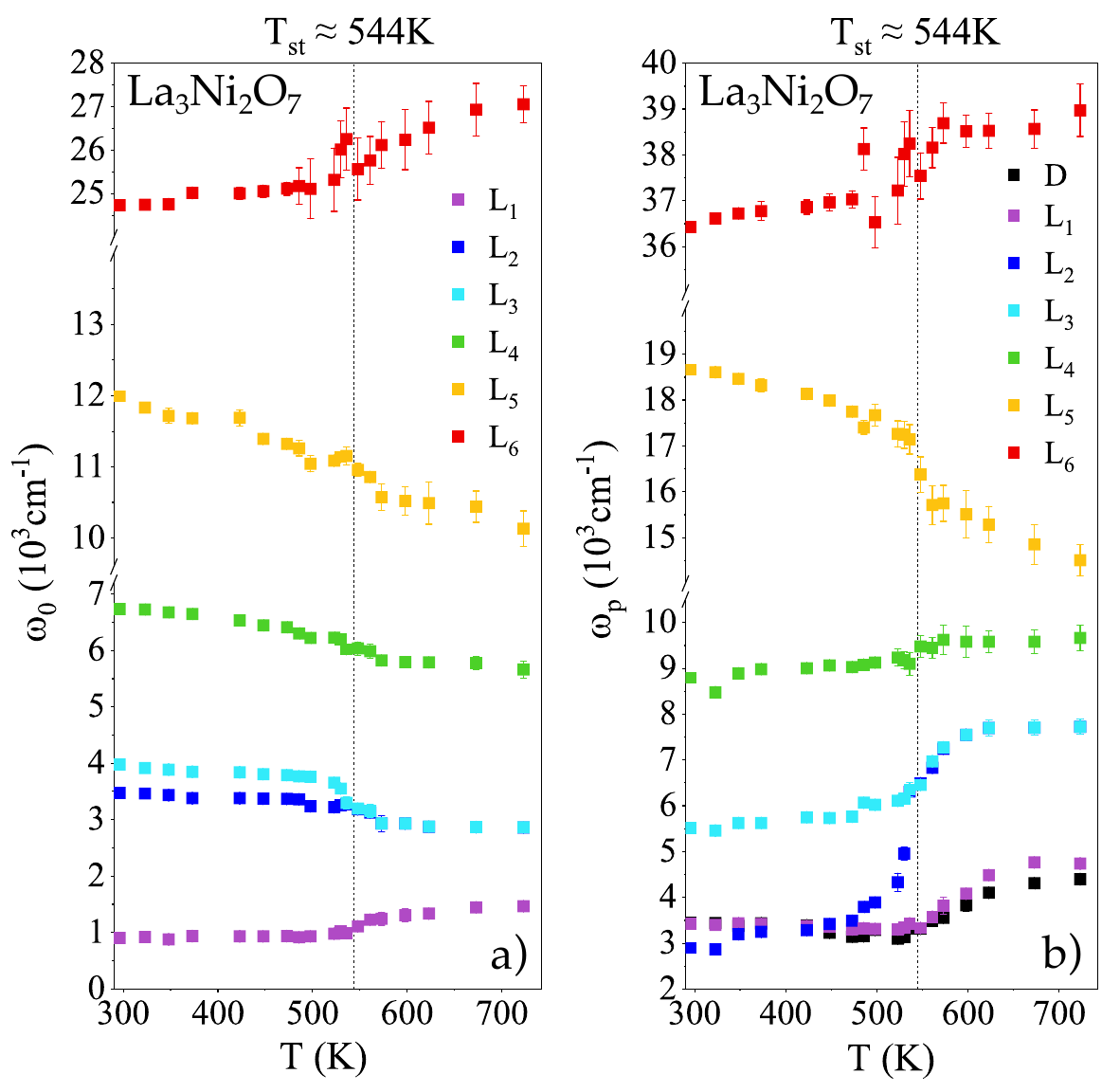}
\caption{\label{fig:fig3} Resonance and plasma frequencies, $\omega_0$ and $\omega_p$, of the Drude $D$ and Lorentzians $L_j$ across $T_{\mathrm{st}}\simeq544$~K. The most pronounced effect of the structural transition on the optical components are observed for $L_2$ and $L_3$ where their positions, plasma frequencies and linewidths (see Fig.~S5) merge accompanied by a strong spectral weight enhancement $K_{\infty,j} \propto \omega_{p,j}^2$ (5 times for $L_2$). $L_5$ undergoes an important loss in spectral weight about 1.7 times across $T_{\mathrm{st}}$.}
\end{figure}

\medskip
\textbf{\textit{Drude--Lorentz analysis}}\ -- To track the optical excitations involved in the spectral-weight redistribution, the reflectivity spectra were analyzed using the \textsc{RefFIT} software~\cite{AKuz05} with a phenomenological Drude--Lorentz model consisting of one Drude term $D$ and six Lorentz oscillators $L_j$, with $\epsilon_\infty=3.5$ fixed for all temperatures (see Fig.~S4 and Methods~\cite{suppmat}). The same model was applied consistently throughout the temperature range to follow the systematic evolution of the optical excitations. Figs.~\ref{fig:fig3}(a,b) show the temperature dependence of the fitted resonance frequencies $\omega_0$ and plasma frequencies $\omega_p$ of the different components. Upon heating, $L_2$--$L_5$ generally shift to lower frequencies, whereas $L_1$ and $L_6$ move to higher frequencies. The fitted plasma frequencies increase predominantly for $L_2$ and $L_3$, while $L_5$ undergoes a pronounced depletion around $T_{\mathrm{st}}$. Notably, the resonance frequencies of $L_2$ and $L_3$ progressively approach each other and become nearly degenerate above $T_{\mathrm{st}}$. Together with the evolution of their linewidths $\Gamma$ shown in Fig.~S5, this behavior reveals a progressive merging of the two finite-energy contributions across the structural transition. 

A component-resolved integration of the fitted optical response being linked to $\omega_{p,j}^2$ (see Fig.~S6 and Methods~\cite{suppmat}) further associates the largest low-energy spectral-weight enhancement with $L_2$ (5 times more) and $L_3$ (2 times more), with a smaller contribution for $L_4$, while spectral-weight losses occur predominantly in $L_5$ (1.7 times less) and $L_6$ within the measured spectral range (from 0 to 26\,000~cm$^{-1}$). The gains in spectral weight associated with $L_1$ and the Drude term $D$ remain comparatively small. Although the decomposition is phenomenological and the individual oscillator strengths are model dependent, the systematic evolution of the finite-energy excitations identifies $L_2$ and $L_3$ as the components most strongly affected across $T_{\mathrm{st}}$. These results provide a phenomenological description of the finite-energy excitations underlying the broadband spectral-weight redistribution established independently in Fig.~\ref{fig:fig2}.

\par
\begin{figure}[htp!]
\includegraphics[scale=0.9]{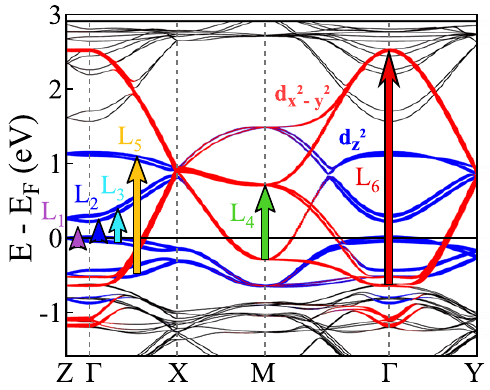}
\caption{\label{fig:fig4} Comparison of our optical responses $L_j$ with band structure calculation at ambient conditions (adapted from \cite{JY24}). We identify our 6 Lorentzians with interband transitions within the states of $3d_{z^2}$ ($L_1$, $L_2$, $L_3$ and $L_5$) and $3d_{x^2-y^2}$ ($L_4$ and $L_6$) orbitals of Ni. Above $T_{\mathrm{st}}\simeq544$~K, $L_2$ and $L_3$ merge (see Fig.~\ref{fig:fig3}) meaning the splitting of the $3d_{z^2}$ antibonding bands disappears within the high-symmetry crystal structure.}
\end{figure}

\medskip
\noindent\textbf{\textit{Link to the band structure}}\ -- To relate the optical excitations to the underlying electronic structure, we compare our results in Fig.~\ref{fig:fig4} with previous orbital-resolved electronic band structure and optical calculations for La$_3$Ni$_2$O$_7$~\cite{Sun23,JY24,BGeisler24,Geisler24}. The states near the Fermi level are predominantly derived from Ni $3d_{z^2}$ and $3d_{x^2-y^2}$ orbitals, with the $3d_{z^2}$ states split by the bilayer coupling into bonding and antibonding bands. While the reported DFT+$U$ calculations place the bonding states below the Fermi level, the low-energy optical excitations resolved in our spectra in Fig.~\ref{fig:fig3} suggest a different relative alignment of these states. 
 
To account for the lowest-energy excitation $L_1$, we consider a level scheme in which the bonding bands are split, with the Fermi level lying between them (see Fig.~\ref{fig:fig4}), and the splitting increasing across the structural transition, as illustrated in Fig.~\ref{fig:fig3}(a) (purple squares). Within this scheme, $L_1$ corresponds to the transition between the two bonding branches around the $\Gamma$ point (purple arrow), while $L_2$ and $L_3$ correspond to transitions from the occupied lower bonding branch to the two unoccupied antibonding bands (dark- and light-blue arrows). Lorentzian $L_6$ with a position about 3~eV can be assigned to the large interband transition in red arrow between the two branches located at the $\Gamma$ point corresponding to the $3d_{x^2-y^2}$ orbital of Ni with the occupied branch giving rise to the $\alpha$ Fermi surface in ARPES~\cite{JY24}. Lorentzians $L_4$ and $L_5$ with positions about 0.8~eV and 1.5~eV could respectively correspond to interband transitions between the two bands of $3d_{x^2-y^2}$ orbital located around the $M$ point in green arrow (lower occupied branch giving the $\beta$ Fermi surface~\cite{JY24}), and other lower and higher bands of $3d_{z^2}$ orbital located along the $\Gamma - X$ direction in orange arrow.

According to band calculations at high pressure~\cite{Sun23,BGeisler24,Geisler24}, the bonding bands are merging and crossing the Fermi level whereas the antibonding branches get a larger splitting. In our situation, at high temperature and ambient pressure, we observe the opposite: 
\begin{itemize}
  \item The position $\omega_0$ of $L_1$ mode increases: the bonding bands get more split.
  \item $L_2$ and $L_3$ merge: antibonding branches become degenerated.
\end{itemize}
This difference between high-pressure and high-temperature regimes could explain the strong deviation on the Drude response: strong enhancement by a reported 100 factor on $\omega_p^2$ at high pressure~\cite{BMichon26} explainable by the fact bonding bands are crossing the Fermi energy, and almost no change at high temperature because bonding branches are driven away from Fermi level while antibonding states are still well above this level. This difference may reflect the distinct lattice responses to pressure and temperature: although both drive the crystal structure toward the same high-symmetry, tilt-free phase, pressure contracts the lattice whereas heating expands it, resulting in different modifications of the electronic band structure.

%This assignment provides a consistent microscopic picture for the three low-energy optical excitations. The evolution of $L_2$ and $L_3$ can then be considered in relation to the structural transition where the splitting within the antibonding branches vanishes above $T_{\mathrm{st}} \approx 544$~K. In La$_3$Ni$_2$O$_7$, the two NiO$_2$ layers within each bilayer are coupled through the inner apical oxygen, which mediates the interaction between Ni $3d_{z^2}$ orbitals. The transition from the tilted to the tilt-free structure modifies the local Ni--O geometry and is expected to affect the $3d_{z^2}$--O $2p_z$ hybridization and interlayer coupling~\cite{Geisler24,BGeisler24}. Within the proposed assignment, the progressive convergence of $L_2$ and $L_3$ across $T_{\mathrm{st}}$, together with their pronounced spectral-weight enhancement, is therefore consistent with a substantial reconstruction involving the bilayer $3d_{z^2}$ orbital sector. More generally, the simultaneous evolution of the excitation energies and spectral weights establishes that the structural transition is accompanied by a pronounced reorganization of the finite-energy electronic response, providing a microscopic basis for the broadband spectral-weight redistribution observed in Fig.~\ref{fig:fig2}.

The evolution of $L_2$ and $L_3$ can therefore be understood as a signature of the reconstruction of the bilayer $3d_{z^2}$ electronic sector across the tilt-free transition. The suppression of tilts in NiO$_6$ octahedra modifies the local Ni–O geometry and, consequently, the $3d_{z^2}$--O$2p_z$ hybridization and apical-oxygen-mediated interlayer coupling. The progressive convergence of $L_2$ and $L_3$, together with their pronounced spectral-weight enhancement, thus provides direct optical evidence for a substantial reconstruction of the bilayer $3d_{z^2}$-derived electronic states. More generally, the simultaneous evolution of excitation energies and spectral weights establishes that the structural transition is accompanied by a pronounced reorganization of the finite-energy electronic response, providing a microscopic basis for the broadband spectral-weight redistribution observed in Fig.~\ref{fig:fig2}.

\par\medskip
\noindent\textbf{\textit{Summary}}\ -- In this work, we investigated the temperature evolution of the electronic response of La$_3$Ni$_2$O$_7$ across the transition from the tilted to the tilt-free structures at $T_{\mathrm{st}}\simeq544$~K using broadband optical spectroscopy at ambient pressure, complemented by HT Raman measurements. The Raman response tracks the accompanying lattice evolution and reveals a coupling between the phonon excitations and the electronic continuum across the structural transition.

Broadband optical spectroscopy reveals a pronounced reorganization of the electronic response across $T_{\mathrm{st}}$. The differential conductivity shows a strong reshaping of the low-energy response through the structural transition, while the integrated spectral weight exhibits a step-like enhancement at low energies accompanied by a depletion at higher energies. Importantly, the low-energy spectral-weight gain does not fully compensate the loss within the measured range. Together with the tendency toward an enhanced response in the UV range, this indicates that the tilt-free transition reorganizes the electronic spectral weight over a remarkably broad energy scale, extending from the low-energy excitations down to 15~meV toward the ultraviolet above 3.2~eV.

Drude–Lorentz analysis further identifies the electronic excitations underlying this redistribution. The changes are dominated by finite-energy excitations rather than the Drude response, with $L_2$ and $L_3$ accounting for the largest spectral-weight gain. Their resonance energies progressively approach each other and the two contributions merge across the structural transition above $T_{\mathrm{st}}\simeq544$~K. Combined with orbital-resolved electronic band structure, these observations point to a reconstruction of the bilayer Ni $3d_{z^2}$-derived states. The suppression of tilts in NiO$_6$ octahedra modifies the Ni–O$_{\mathrm{ap}}$–Ni geometry and, consequently, the $3d_{z^2}$--O$2p_z$ hybridization and apical-oxygen-mediated interlayer coupling.

This reconstruction is particularly relevant in light of theoretical descriptions in which the $3d_{z^2}$ orbital sector plays a central role in the magnetic and superconducting properties of La$_3$Ni$_2$O$_7$~\cite{YH26}. In orbital-selective models, the interlayer coupling of the $3d_{z^2}$ states generates an interlayer exchange $J_\perp$, whose enhancement suppresses magnetic order and favors interlayer pairing. Our observation that the tilt-free transition strongly reconstructs the same orbital sector therefore provides an experimental link between the lattice distortion, the electronic structure, and the interlayer interactions implicated in superconductivity. Although the temperature- and pressure-driven transitions are not electronically identical, their common suppression of the octahedral tilts suggests that the lattice provides a direct handle on the $3d_{z^2}$-derived electronic degrees of freedom relevant to the high-$T_c$ superconducting state. The ambient-pressure transition thus provides a reference point for understanding how suppression of the tilts reorganizes the electronic structure toward the pressure-induced superconducting regime.

\par\medskip
\noindent\textbf{\textit{Acknowledgments}}\ -- Spectroscopy experiments presented above were performed at the SMIS infrared spectromicroscopy beamline of the SOLEIL synchrotron. The work at IOPCAS is supported by the National Key R\&D Program of China (Grant No. 2023YFA1406100), the National Natural Science Foundation of China (Grant Nos. 12522407, 12494592, and 12025408).

B.M. coordinated the project and defined the project objectives. M.K. and B.M. carried out the high-temperature infrared reflectivity measurements with assistance from G.N. and F.B.. B.M., G.N. and F.B. conducted complementary high-temperature Raman measurements. M.K. performed the analysis and carried out the interpretation of the entire dataset with feedback from B.M.. M.K. and B.M. wrote the manuscript, incorporating comments from the coauthors. Y.Y., B.W., J.S. and J.C. synthesized and characterized the La$_3$Ni$_2$O$_7$ single crystals. F.B. provided postdoctoral funding for B.M.

%% file: Supplementary.tex
\clearpage

\section*{\label{sec:Metho}Section 1. Methods}

\subsection{Single-crystal growth}
Single crystals of La$_3$Ni$_2$O$_7$ were grown by a molten salt flux evaporating method at ambient pressure, which follows the procedure as described in Ref.~\cite{FLi26}. They are platelets with a square-shape (ab plane) of about 40-50~$\mu$m size and with a thickness around 10-30~$\mu$m (along the c-axis). High-purity La$_2$O$_3$ (99.99\%, dried at 1000~\textdegree{C} overnight) and NiO were used as starting materials. The oxide powders were weighed and grounded, and then mixed with anhydrous K$_2$CO$_3$ as flux in a solute-to-flux mass ratio of 1:15. To prevent moisture absorption, all the processes were performed inside a glovebox. The mixture was placed in an alumina crucible that was covered with a lid to control the evaporation rate. The crystals were grown in the furnace at a temperature of 1000-1050~\textdegree{C} for 72~h with evaporating the flux gradually and then were cooled to room temperature naturally. The crystals were extracted by soaking the mixture in deionized water. To reduce the oxygen vacancies and further improve the stoichiometry, we annealed the as-grown crystals in a tube furnace under a flowing oxygen atmosphere at 500~\textdegree{C} for 5 days, followed by furnace cooling to room temperature.

\subsection{Experimental setups}

For the high-temperature, ambient-pressure measurements, the sample was mounted on an optical heating stage (Linkam) under a dry N$_2$ gas flow atmosphere, allowing the temperature to be controlled across the structural transition up to 450~\textdegree C to prevent sample deterioration. Temperature-dependent reflectivity measurements were performed on the $ab$ plane of La$_3$Ni$_2$O$_7$ over the 296--723~K temperature range using unpolarized light and three complementary spectroscopic setups covering the far-infrared (FIR), mid- and near-infrared (MIR--NIR), and near-infrared to visible (NIR--VIS) spectral ranges.

The synchrotron-based FIR measurements were performed using a Thermo Scientific iS50 spectrometer equipped with a Si beamsplitter, coupled to a Thermo Nicolet Nicplan infrared microscope and an IR Labs Si bolometer. The MIR--NIR measurements were performed using a Thermo Scientific 5700 spectrometer equipped with KBr and CaF$_2$ beamsplitters for the MIR and NIR ranges, respectively, and coupled to a Thermo Scientific Continuµm XL infrared microscope equipped with a MCT/A detector and Globar and white-light sources.

The NIR--VIS reflectivity measurements covered approximately 8300--30,000~cm$^{-1}$ and were performed using a home-built NIR--VIS microscope. A broadband NIR--visible lamp was used for illumination, and the reflected light was collected with a 20$\times$ Thorlabs mirror objective. The reflected light was analyzed using an Andor KY328i spectrometer equipped with a DU420\_BVF CCD detector and a 299.972~lines/mm grating blazed at 500~nm. Spectra were acquired with an exposure time of 0.2~s and 100 accumulations.

The partial spectral overlap between adjacent ranges was used to construct continuous broadband reflectivity spectra. For the optical analysis presented in the main text, the spectra were restricted to energies below 26,000~cm$^{-1}$ because of reduced reproducibility and increased noise at higher frequencies.

For each spectral range, a reference spectrum was acquired immediately before the corresponding sample spectrum at each temperature, with both measurements performed under identical optical conditions. A gold-flake reference was used for the FIR and MIR ranges, whereas an aluminum-foil reference was used for the NIR--VIS range. The reference-sample sequence, enabling to get the raw reflectivity data by the ratio $R(\omega,T) = I_{\mathrm{samp}}(\omega,T)/I_{\mathrm{ref}}(\omega,T)$, was repeated independently at every temperature and for each spectral range, thereby minimizing systematic time variations in the experimental response during the temperature-dependent measurements.

Complementary high-temperature Raman measurements were performed at ambient pressure using a 532~nm laser, with the sample mounted on an optical heating stage (Linkam). Raman spectra were acquired upon heating across the structural transition at $T_{\mathrm{st}}\simeq544$~K in order to monitor the accompanying evolution of the lattice dynamics.

\subsection{Data analysis and fitting}

The temperature-dependent reflectivity spectra were analyzed using the \textsc{RefFIT} software~\cite{AKuz05} with a phenomenological Drude--Lorentz model consisting of one Drude term and six Lorentz oscillators. The same model was applied consistently over the entire temperature range in order to track the systematic evolution of the optical excitations, with the high-frequency dielectric constant fixed to $\varepsilon_\infty=3.5$ for all temperatures. The complex dielectric function was described as
\begin{equation} \label{eq1}
\varepsilon(\omega)=\varepsilon_\infty+\sum_j\frac{\omega_{p,j}^2}{\omega_{0,j}^2-\omega^2-i\Gamma_j\omega},
\end{equation}
where $\omega_{p,j}$, $\omega_{0,j}$, and $\Gamma_j$ denote the plasma frequency, resonance frequency, and linewidth of the $j$-th oscillator, respectively. The free-carrier response corresponds to $\omega_{0,j}=0$ and is referred to as the Drude term $D$, whereas the six finite-frequency oscillators are denoted $L_1$--$L_6$.

At ambient pressure and near-normal incidence, the calculated reflectivity is related to the complex dielectric function through the complex refractive index $\tilde{n}(\omega)=\sqrt{\varepsilon(\omega)}$ according to
\begin{equation} \label{eq2}
R(\omega)=\left|\frac{\tilde{n}(\omega)-1}{\tilde{n}(\omega)+1}\right|^2.
\end{equation}

The quality of the reflectivity fits was evaluated by comparing the measured spectra with the corresponding fitted spectra over the common spectral range.

The temperature dependence of the fitted resonance frequencies $\omega_0$ and plasma frequencies $\omega_p$ is presented in Figs.~3(a,b) (main text), while the corresponding linewidths $\Gamma$ of the Drude and Lorentz components are shown in Figs.~\ref{fig:figS5}(a,b).

The fitting of reflectivity data were subsequently used to extrapolate the measured reflectivity down to 0 and up to 1\,000\,000~cm$^{-1}$. These extrapolated reflectivity data are used to extract the complex optical conductivity $\tilde{\sigma}(\omega)=\sigma_1(\omega)+i\sigma_2(\omega)$ thanks to a Kramers-Kronig transformation. The real part $\sigma_1(\omega)$ and its temperature-induced variation are discussed in Figs.~1(b),2(a) (main text), whereas the corresponding imaginary part $\sigma_2(\omega)$ and its differential response $\Delta\sigma_2(\omega,T)=\sigma_2(\omega,T)-\sigma_2(\omega,296~\mathrm{K})$ are shown in Figs.~\ref{fig:figS1}(a,b).

\subsection{Spectral-weight analysis}

To quantify the temperature-induced redistribution of the optical spectral weight, the differential optical conductivity was defined relative to the 296~K spectrum as $\Delta\sigma_1(\omega,T)=\sigma_1(\omega,T)-\sigma_1(\omega,296~\mathrm{K})$. The corresponding spectral-weight variation over a frequency interval $[\omega_1,\omega_2]$ was calculated as
\begin{equation} \label{eq3}
\Delta K_{\mathrm{int}}([\omega_1,\omega_2],T)=\frac{2d_c\hbar^2}{\pi e^2}\int_{\omega_1}^{\omega_2}\Delta\sigma_1(\omega,T)\,d\omega,
\end{equation}
where $d_c=5.1$~\AA\ is the mean interlayer spacing in La$_3$Ni$_2$O$_7$. In the main text, the integration ranges 0--8000, 8000--26\,000, and 0--26\,000~cm$^{-1}$ are used to compare the low-energy spectral-weight gain with the higher-energy depletion. The boundary at 8000~cm$^{-1}$ approximately separates these two regimes and lies close to the crossing region observed in the reflectivity spectra $R(\omega,T)$ (Fig.~1(a) in main text).

To continuously follow the redistribution as a function of energy, the cumulative spectral weight was evaluated by progressively increasing the upper integration limit. The absolute cumulative spectral weight $K_{\mathrm{int}}(\omega,T)=K_{\mathrm{int}}([0,\omega],T)$ and its variation relative to 296~K, $\Delta K_{\mathrm{int}}(\omega,T) = K_{\mathrm{int}}(\omega,T) - K_{\mathrm{int}}(\omega,296~K)$, are shown in Figs.~\ref{fig:figS2}(a,b). The differential representation directly reveals how the low-energy spectral-weight gain is progressively surpassed by the depletion occurring at higher energies within the measured spectral range (no compensation up to 26\,000~cm$^{-1}$).

A complementary frequency-resolved analysis was performed by dividing the measured range into successive 1000~cm$^{-1}$ intervals up to 10\,000~cm$^{-1}$ and 2000~cm$^{-1}$ intervals above 10\,000~cm$^{-1}$, as shown in Fig.~\ref{fig:figS3}. Finer intervals are used at low energies to accurately capture the pronounced redistribution associated with the low-energy excitations, whereas coarser intervals suffice at higher energies, where the optical response is dominated by the broad $L_5$ and $L_6$ contributions and varies smoothly with frequency. This analysis provides a more detailed view of the energy ranges contributing to the redistribution and verifies that the overall behavior does not depend on the particular choice of 8000~cm$^{-1}$ as the boundary used in Fig.~2 (main text).

Finally, the spectral weight associated with each fitted optical component was evaluated separately for the Drude term $D$ and the Lorentz oscillators $L_1$--$L_6$ by integrating the isolated contributions of each oscillator in $\sigma_1(\omega)$, $\sigma_{1,j}(\omega)$. For each component $j$: 
\begin{equation} \label{eq4}
K_{\mathrm{int},j}(T) = K_{\mathrm{int},j}([0,26\,000~cm^{-1}],T),
\end{equation}
denotes the spectral weight integrated over the experimentally investigated range from 0 up to 26\,000~cm$^{-1}$.

Similarly, we define: 
\begin{equation} \label{eq5}
K_{\infty,j}(T) = K_{\mathrm{int},j}([0,+\infty],T) = \frac{d_c\hbar^2\varepsilon_0}{e^2}\,\omega_{p,j}^2,
\end{equation}
which denotes the corresponding total spectral weight enclosed inside each mode (Drude and Lorentzians) directly related to the plasma frequency squared $\omega_{p,j}^2$, and where $d_c=5.1$~\AA\ is the mean interlayer spacing in La$_3$Ni$_2$O$_7$ and $\varepsilon_0$ the vacuum permittivity. 

Their respective temperature-induced variations relative to 296~K were defined as $\Delta K_{\mathrm{int},j}(T)=K_{\mathrm{int},j}(T)-K_{\mathrm{int},j}(296~\mathrm{K})$ and $\Delta K_{\infty,j}(T)=K_{\infty,j}(T)-K_{\infty,j}(296~\mathrm{K})$. The resulting component-resolved quantities are shown in Figs.~\ref{fig:figS6}(a--d) and provide a complementary decomposition of the broadband spectral-weight redistribution in terms of the individual fitted optical excitations.

%\subsection{DFT+$U$ band structure calculations}

\FloatBarrier
\bibliography{biblio}

\onecolumngrid

\newpage
\section*{\label{sec:Add}Section 2. Additional data and analysis}

\setcounter{figure}{0}
\renewcommand{\thefigure}{S\arabic{figure}}

\begin{figure*}[h]
\centering
\begin{minipage}{.5\textwidth}
  \centering
  \begin{overpic}[scale=0.4]{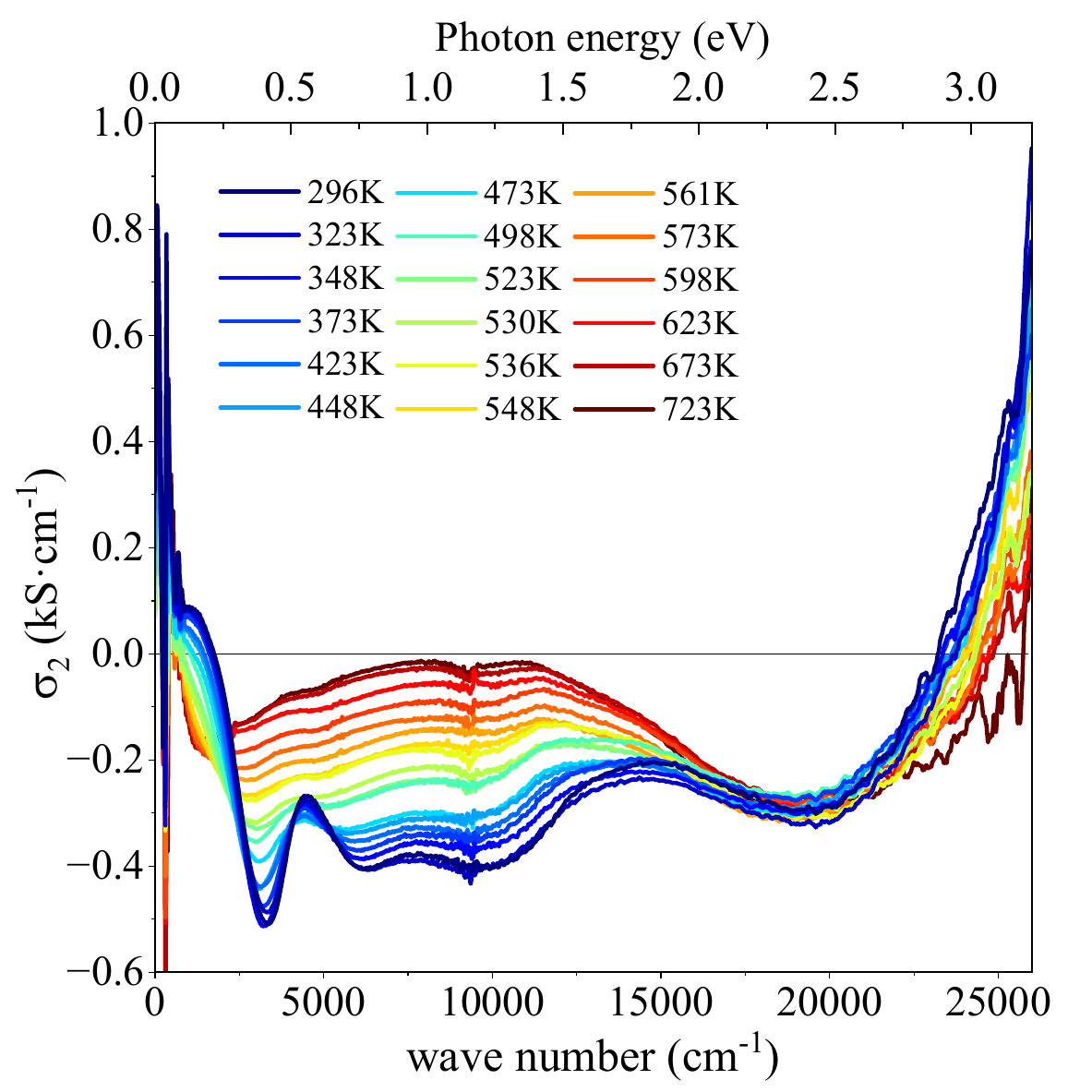}
\put(2,90){\textbf{a)}}
\end{overpic}
\end{minipage}%
\begin{minipage}{.5\textwidth}
  \centering
  \begin{overpic}[scale=0.4]{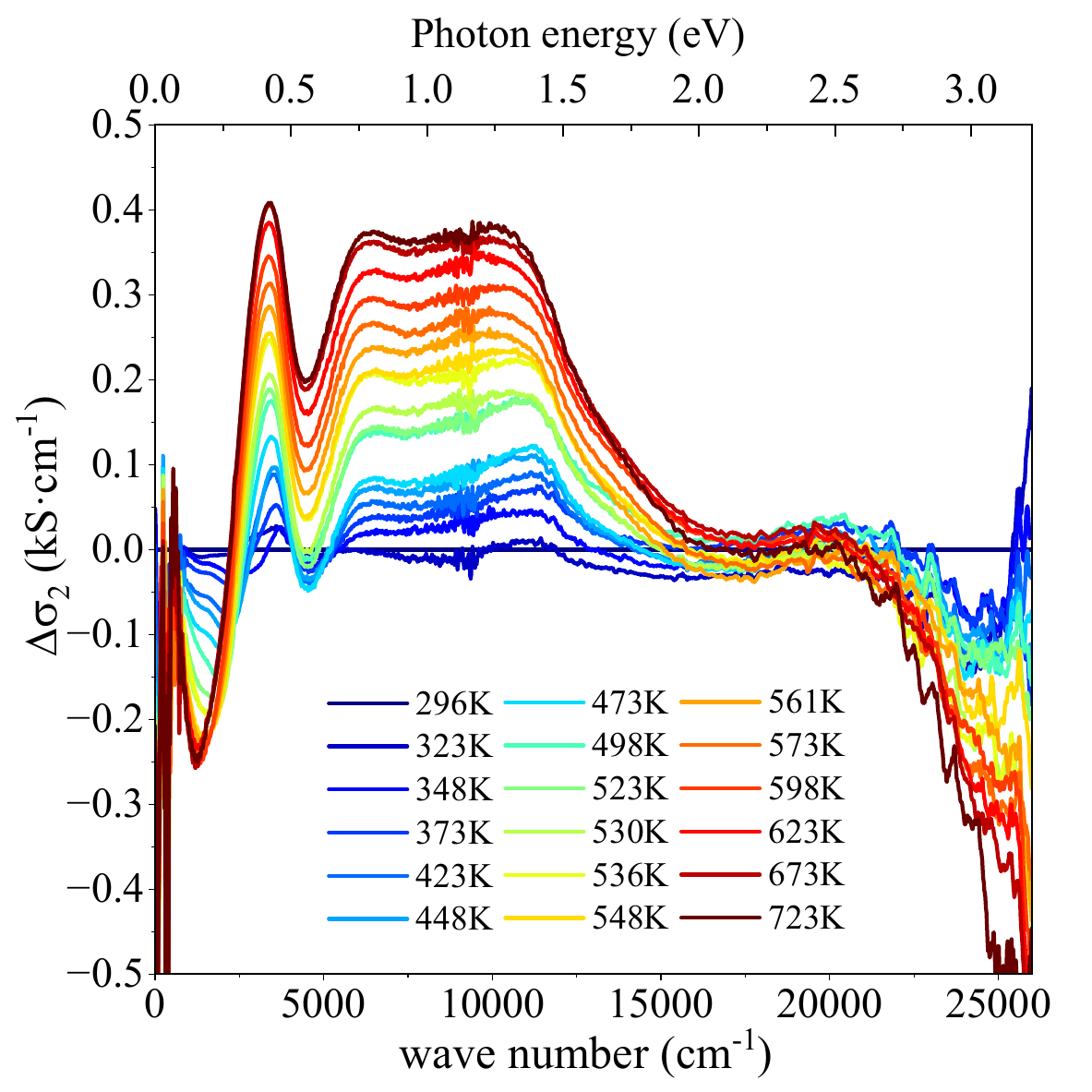}
\put(2,90){\textbf{b)}}
\end{overpic}
\end{minipage}
\caption{\label{fig:figS1} Temperature dependence of $\sigma_2(\omega)$ (imaginary part of the optical conductivity). (a) $\sigma_2(\omega)$ obtained from the Kramers-Kronig analysis of the measured reflectivity $R(\omega)$ at different temperature. (b) Differential imaginary conductivity $\Delta\sigma_2(\omega,T)=\sigma_2(\omega,T)-\sigma_2(\omega,296~\mathrm{K})$ relative to room temperature.}
\end{figure*}

\begin{figure*}[h]
\centering
\begin{minipage}{.5\textwidth}
  \centering
  \begin{overpic}[scale=0.4]{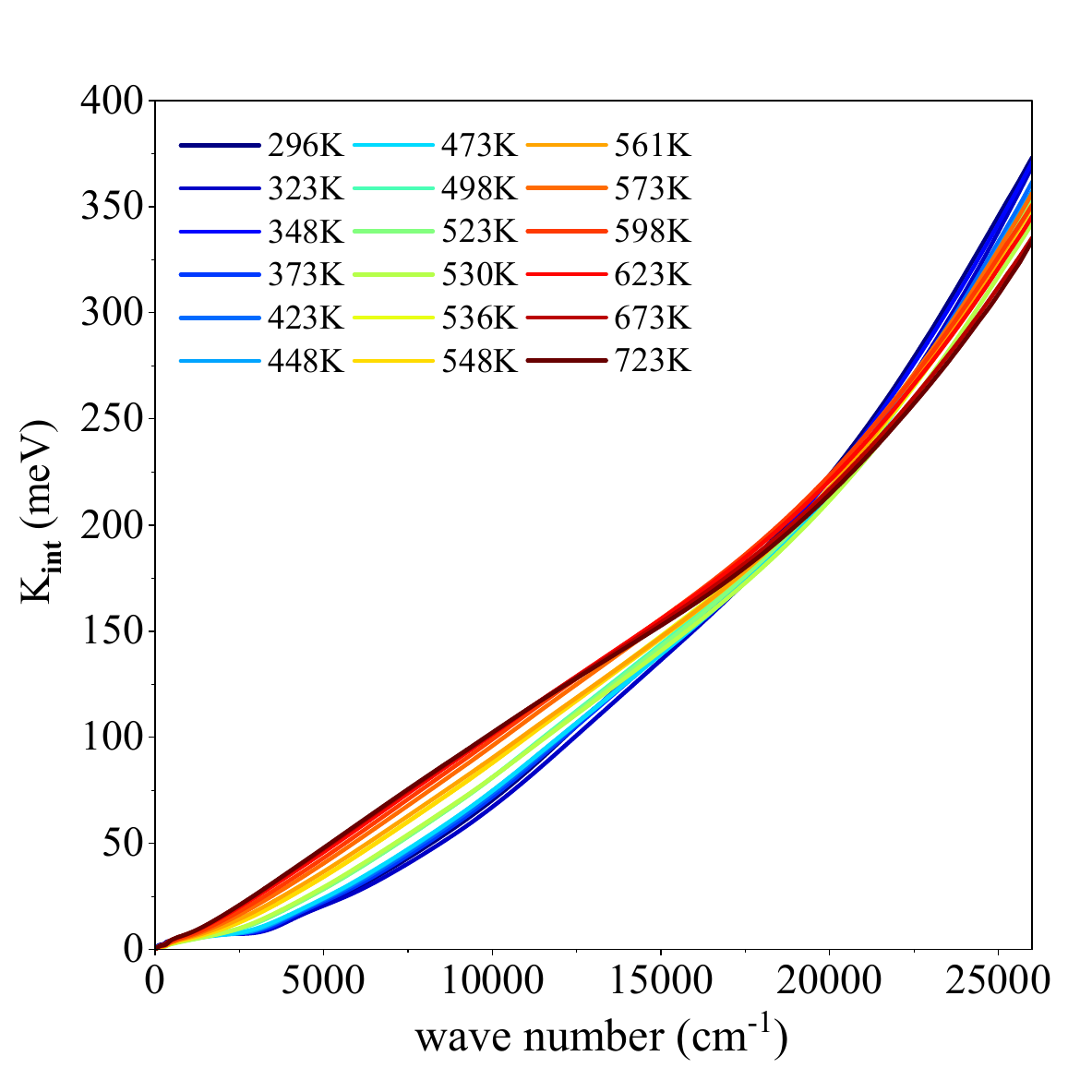}
\put(2,90){\textbf{a)}}
\end{overpic}
\end{minipage}%
\begin{minipage}{.5\textwidth}
  \centering
  \begin{overpic}[scale=0.4]{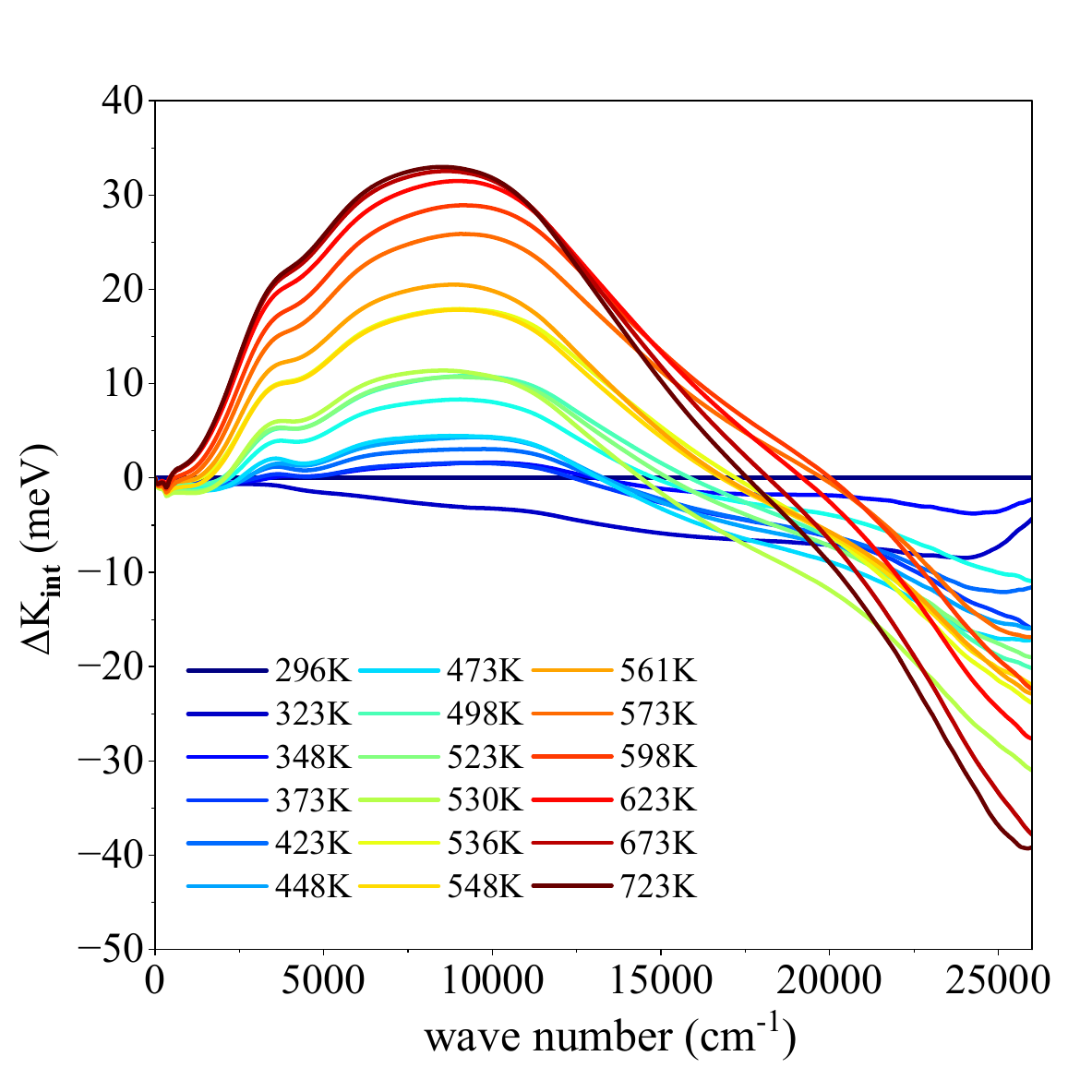}
\put(2,90){\textbf{b)}}
\end{overpic}
\end{minipage}
\caption{\label{fig:figS2} Cumulative optical spectral weight as a function of the upper integration frequency $K_{\mathrm{int}}(\omega,T) = K_{\mathrm{int}}([0,\omega],T)$ at different temperature (see definition in equation~(\ref{eq3})). (a) Integrated spectral weight $K_{\mathrm{int}}(\omega,T)$ obtained by integrating $\sigma_1(\omega,T)$ from zero frequency up to $\omega$. (b) Corresponding differential cumulative spectral weight $\Delta K_{\mathrm{int}}(\omega,T)=K_{\mathrm{int}}(\omega,T)-K_{\mathrm{int}}(\omega,296~\mathrm{K})$.}
\end{figure*}

\begin{figure*}[h]
\centering
\includegraphics[scale=0.4]{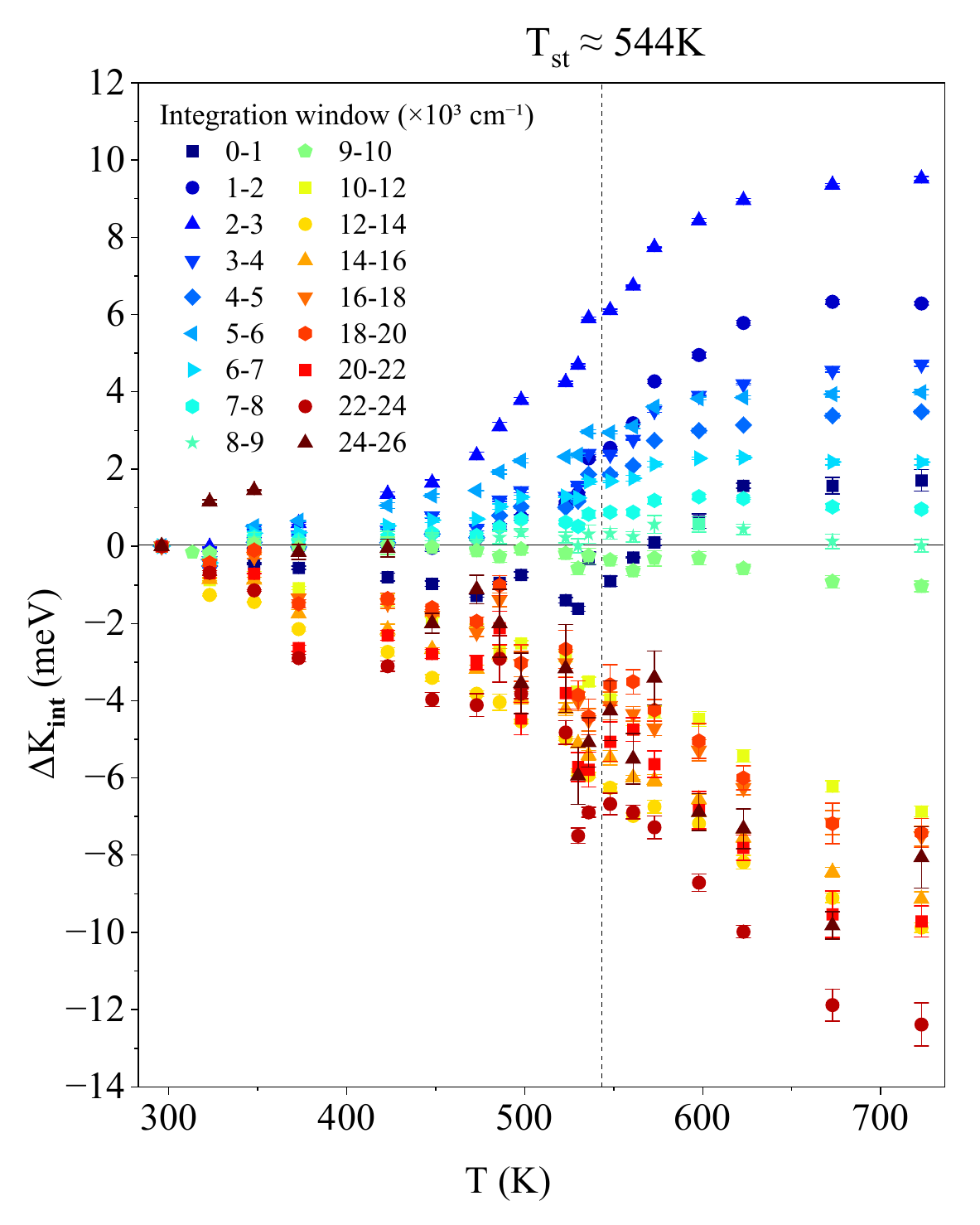}
\caption{\label{fig:figS3} Temperature dependence of the spectral-weight variation $K_{\mathrm{int}}([\omega_1,\omega_2],T)$ evaluated over successive frequency intervals $[\omega_1,\omega_2]$ (see definition in equation~(\ref{eq3})). The spectrum was divided into 1000~cm$^{-1}$ intervals below 10\,000~cm$^{-1}$ and 2000~cm$^{-1}$ intervals above 10\,000~cm$^{-1}$. The vertical dashed line marks the structural transition at $T_{\mathrm{st}}\simeq544$~K. This frequency-resolved analysis highlights the redistribution of spectral weight across multiple energy ranges and shows that the overall evolution is not determined by the specific choice of 8000~cm$^{-1}$ as the boundary used in Figs.~2(a,b) (main text).}
\end{figure*}

\begin{figure*}
\centering
\begin{minipage}{0.4\textwidth}
  \centering
  \begin{overpic}[width=\textwidth]{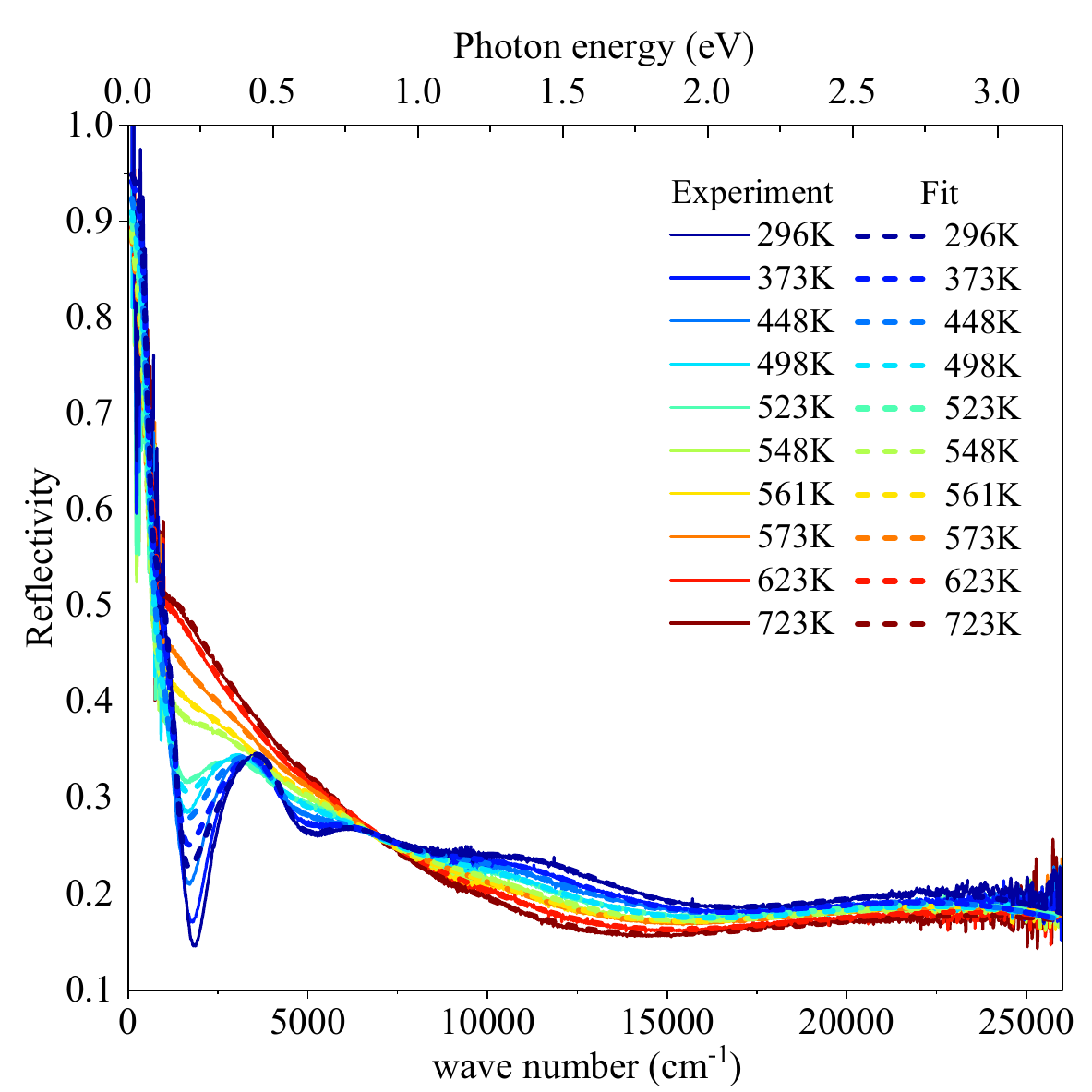}
  \put(2,92){\textbf{a)}}
  \end{overpic}
\end{minipage}

\vspace{0.5cm}

\begin{minipage}{.4\textwidth}
  \centering
  \begin{overpic}[width=\textwidth]{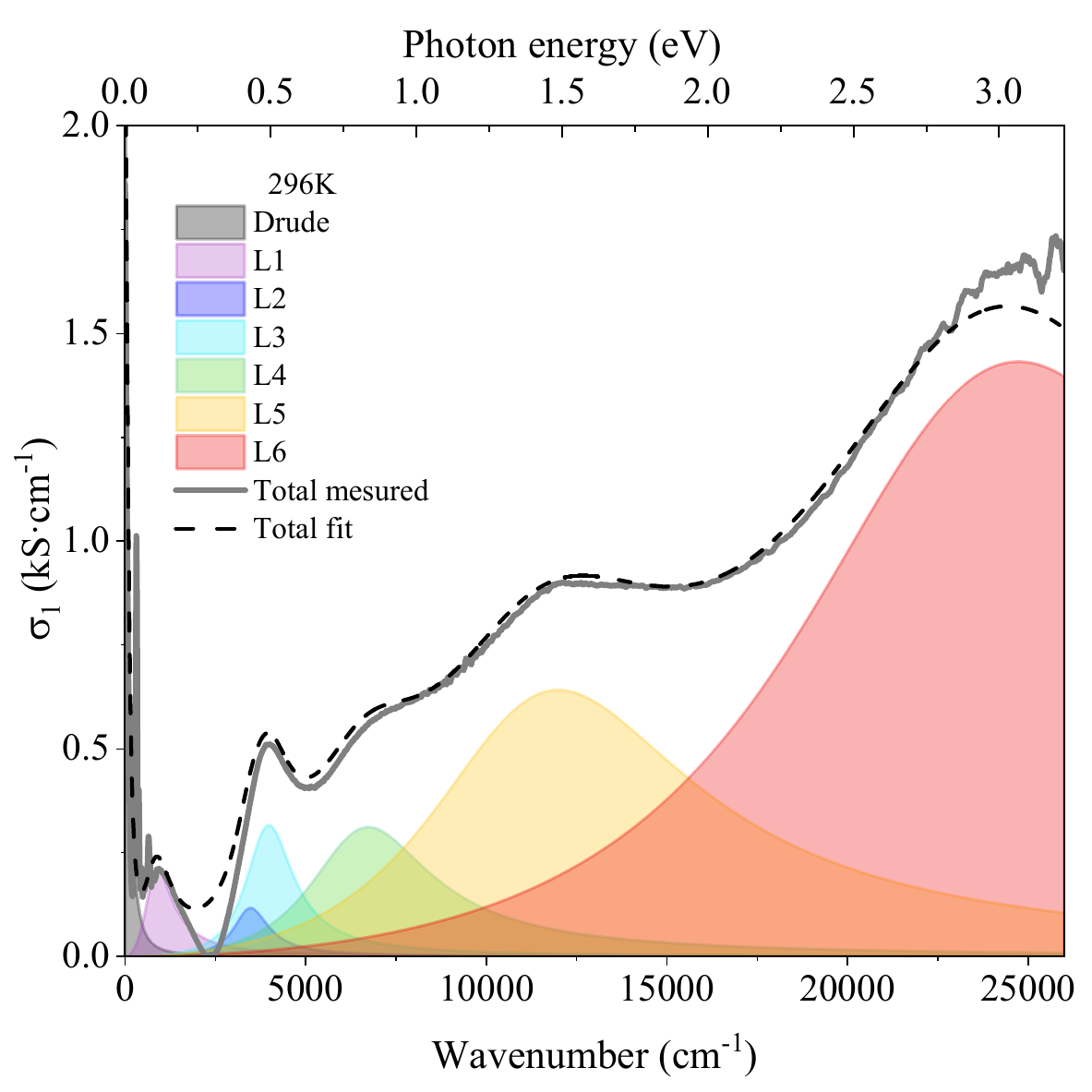}
  \put(2,92){\textbf{b)}}
  \end{overpic}
\end{minipage}%
\begin{minipage}{.4\textwidth}
  \centering
  \begin{overpic}[width=\textwidth]{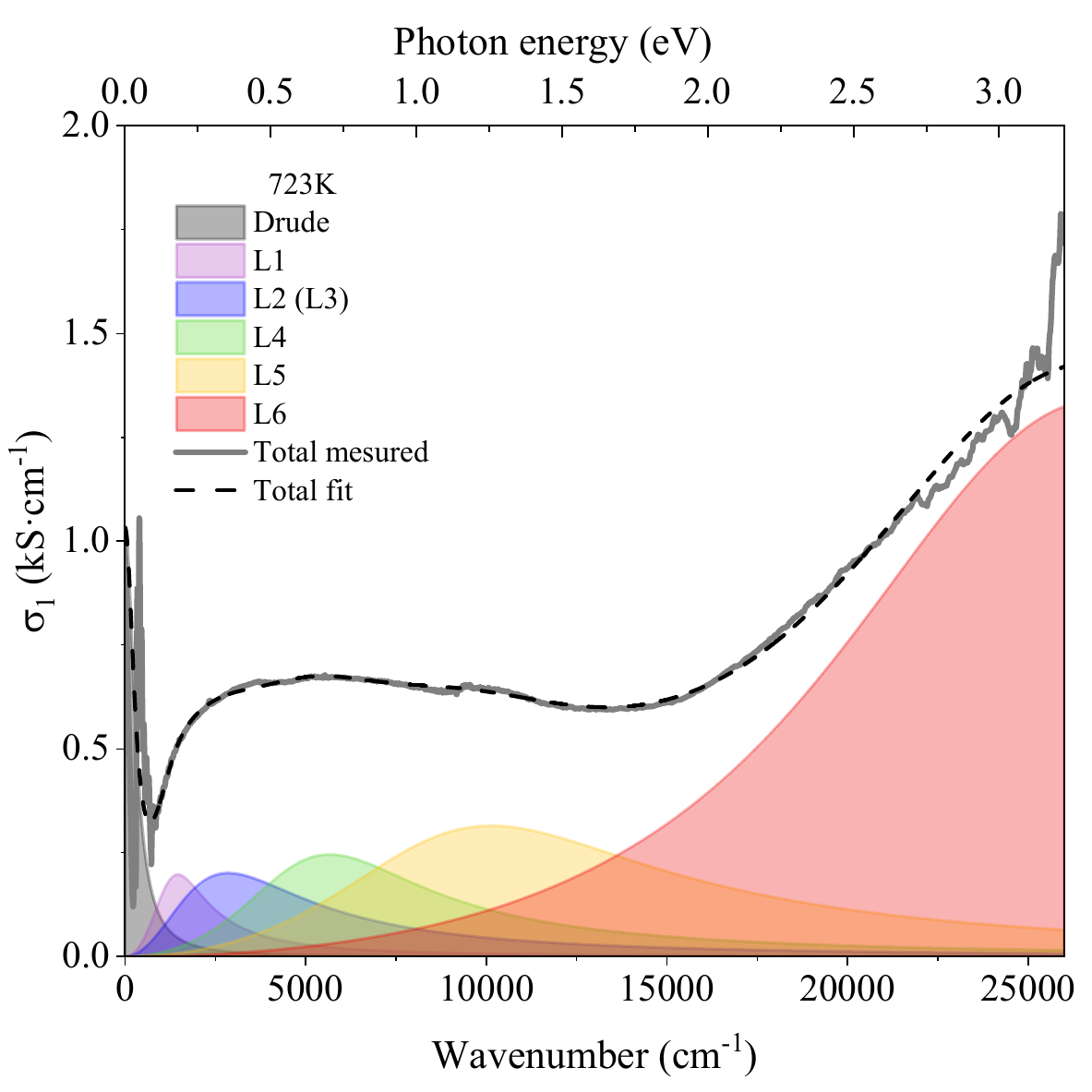}
  \put(2,92){\textbf{c)}}
  \end{overpic}
\end{minipage}

\caption{\label{fig:figS4}
Assessment of the Drude--Lorentz fitting quality.
(a) Measured reflectivity spectra $R(\omega)$ and corresponding fits at selected temperatures.
(b,c) Drude--Lorentz decomposition of the optical conductivity $\sigma_1(\omega)$ at
(b) 296~K and (c) 723~K.
The colored areas represent the individual Drude and Lorentz contributions,
while the solid gray and dashed black curves correspond to the raw
$\sigma_1(\omega)$ and its total fit, respectively. At 723~K, $L_2$ and $L_3$ contributions are identical: $\omega_0$, $\omega_p$ and $\Gamma$ parameters converge to the same values (see Figs.~3(a,b),~\ref{fig:figS5}(a)).
}
\end{figure*}

\begin{figure*}
\centering
\begin{minipage}{.5\textwidth}
  \centering
  \begin{overpic}[scale=0.6]{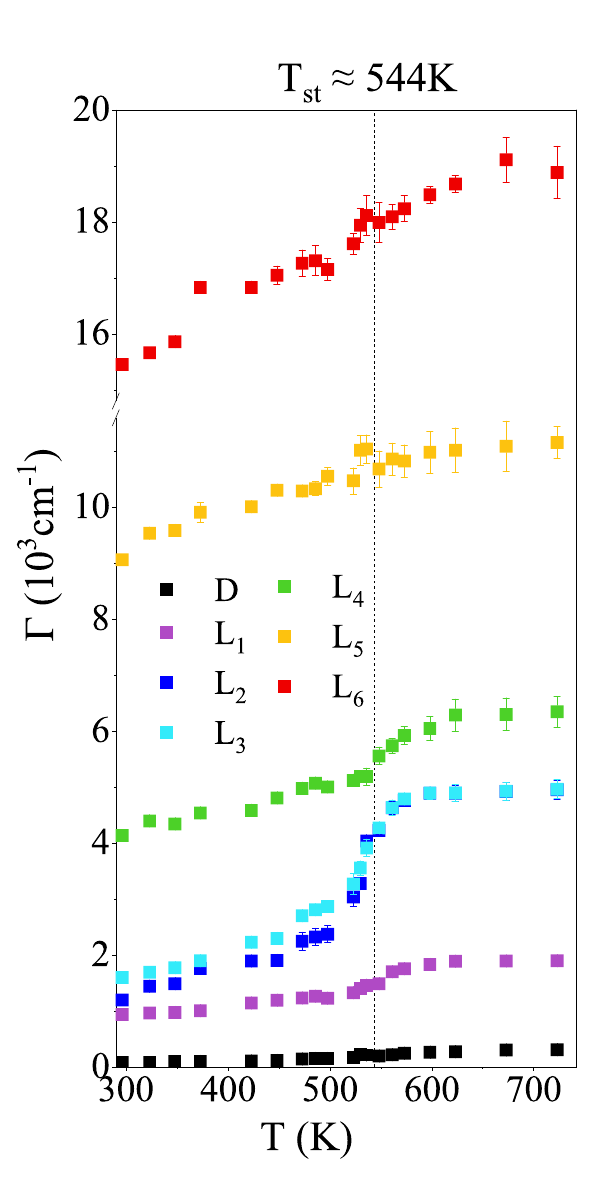}
\put(2,90){\textbf{a)}}
\end{overpic}
\end{minipage}%
\begin{minipage}{.5\textwidth}
  \centering
  \begin{overpic}[scale=0.4]{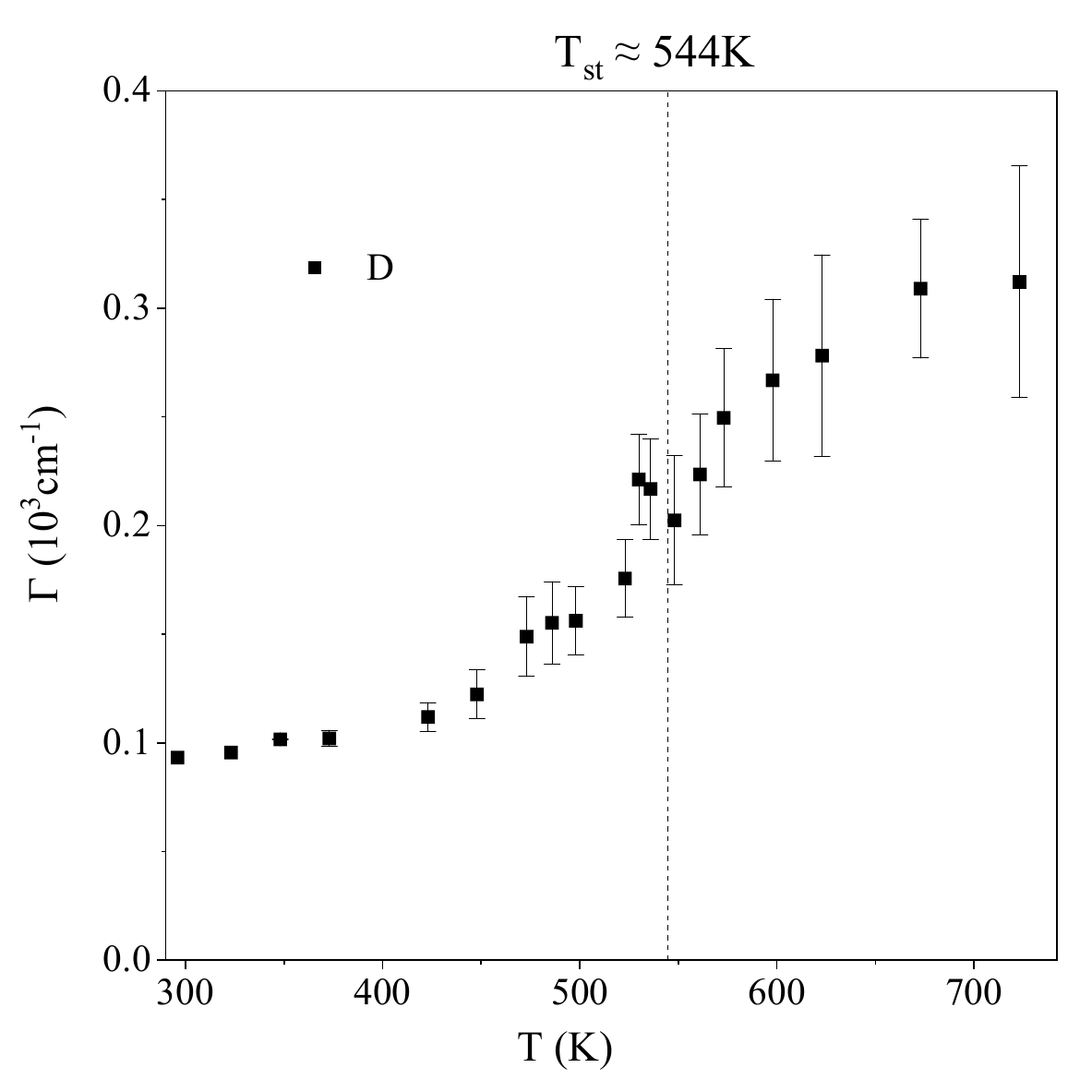}
\put(2,90){\textbf{b)}}
\end{overpic}
\end{minipage}
\caption{\label{fig:figS5} Temperature dependence of the fitted linewidths $\Gamma$. (a) Linewidths of the Drude term $D$ and the Lorentz oscillators $L_1$--$L_6$. (b) Linewidth $\Gamma_D$ of the Drude term shown separately for clarity. The vertical dashed lines mark the structural transition at $T_{\mathrm{st}}\simeq544$~K.}
\end{figure*}

\begin{figure*}
\centering
\begin{minipage}{.5\textwidth}
\centering
\begin{overpic}[scale=0.6]{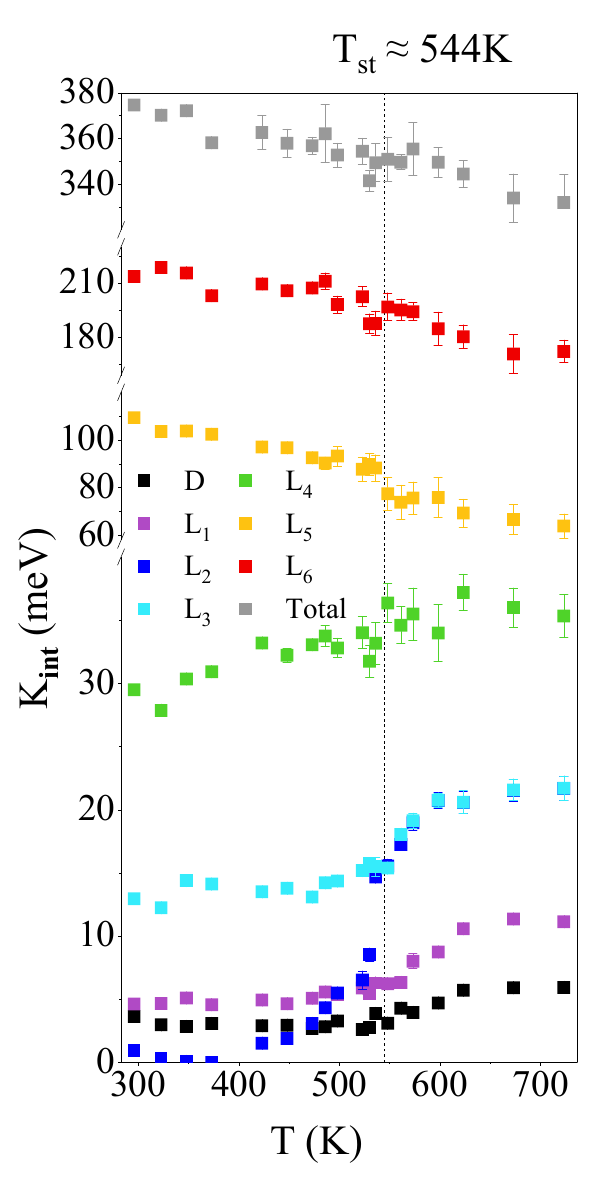}
\put(2,90){\textbf{a)}}
\end{overpic}
\end{minipage}%
\begin{minipage}{.5\textwidth}
\centering
\begin{overpic}[scale=0.6]{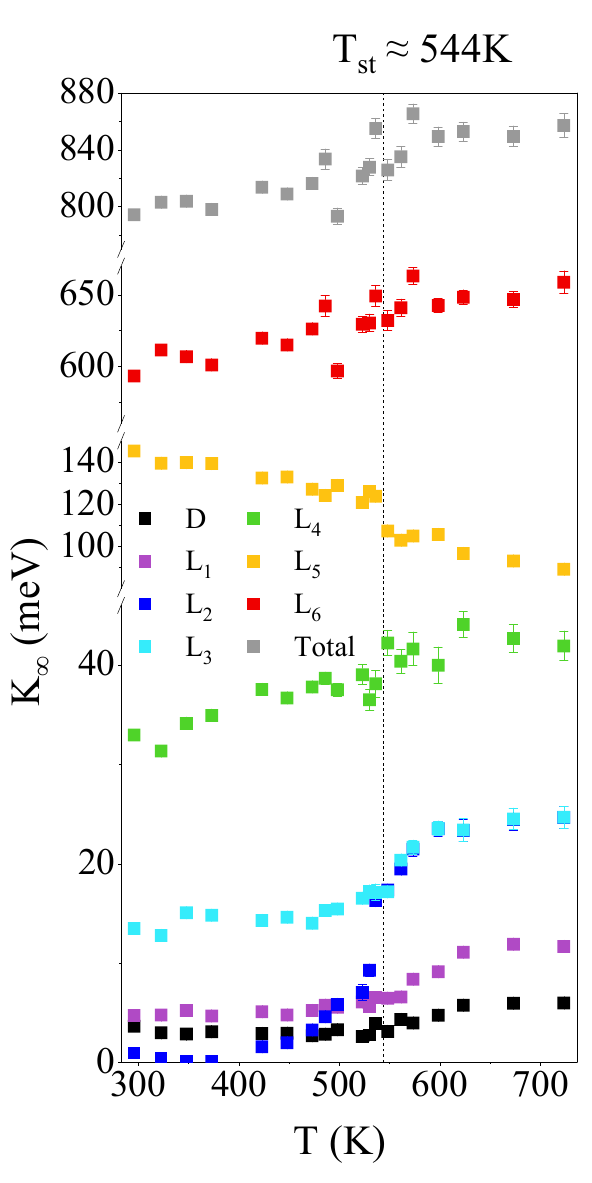}
\put(2,90){\textbf{b)}}
\end{overpic}
\end{minipage}

\begin{minipage}{.5\textwidth}
\centering
\begin{overpic}[scale=0.4]{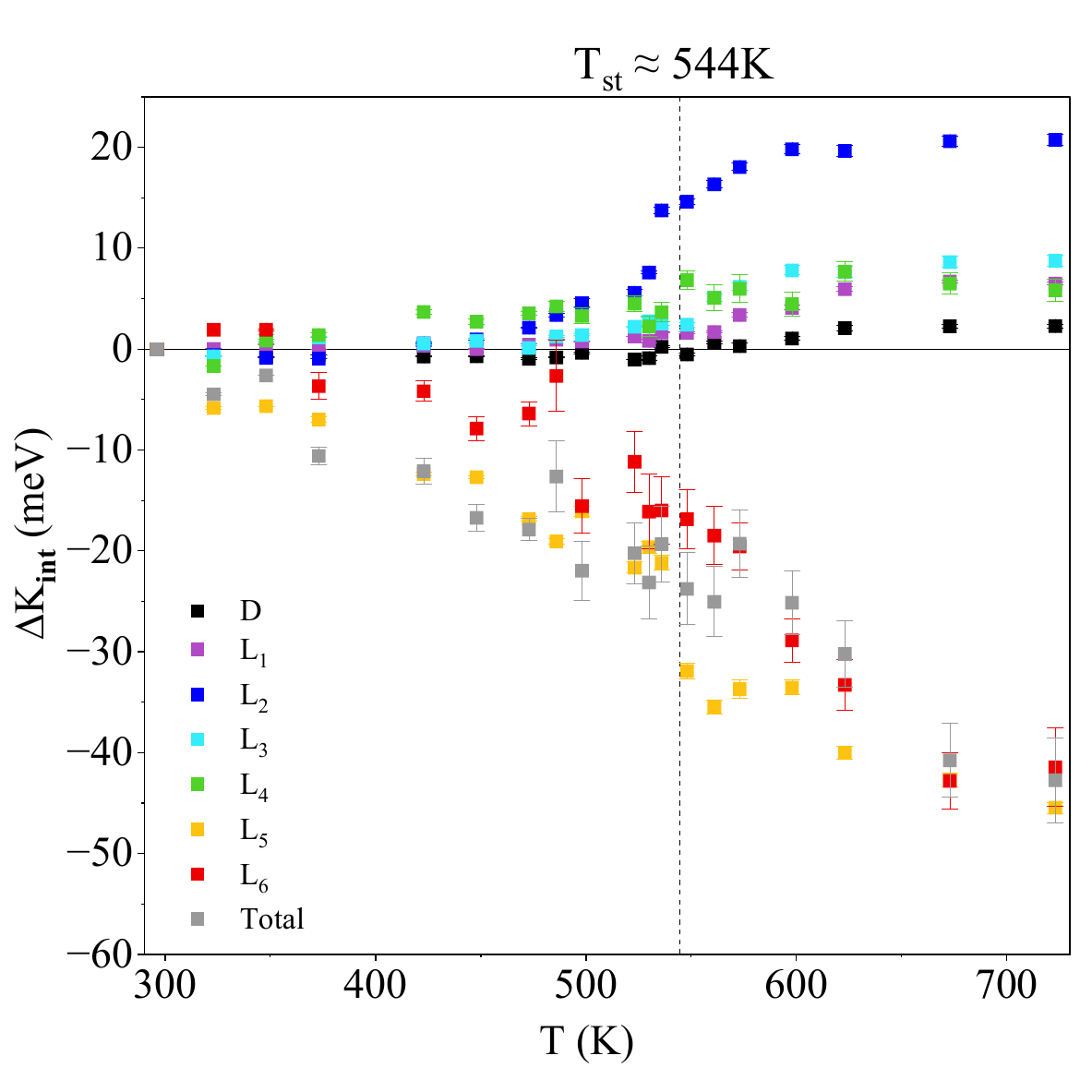}
\put(2,90){\textbf{c)}}
\end{overpic}
\end{minipage}%
\begin{minipage}{.5\textwidth}
\centering
\begin{overpic}[scale=0.4]{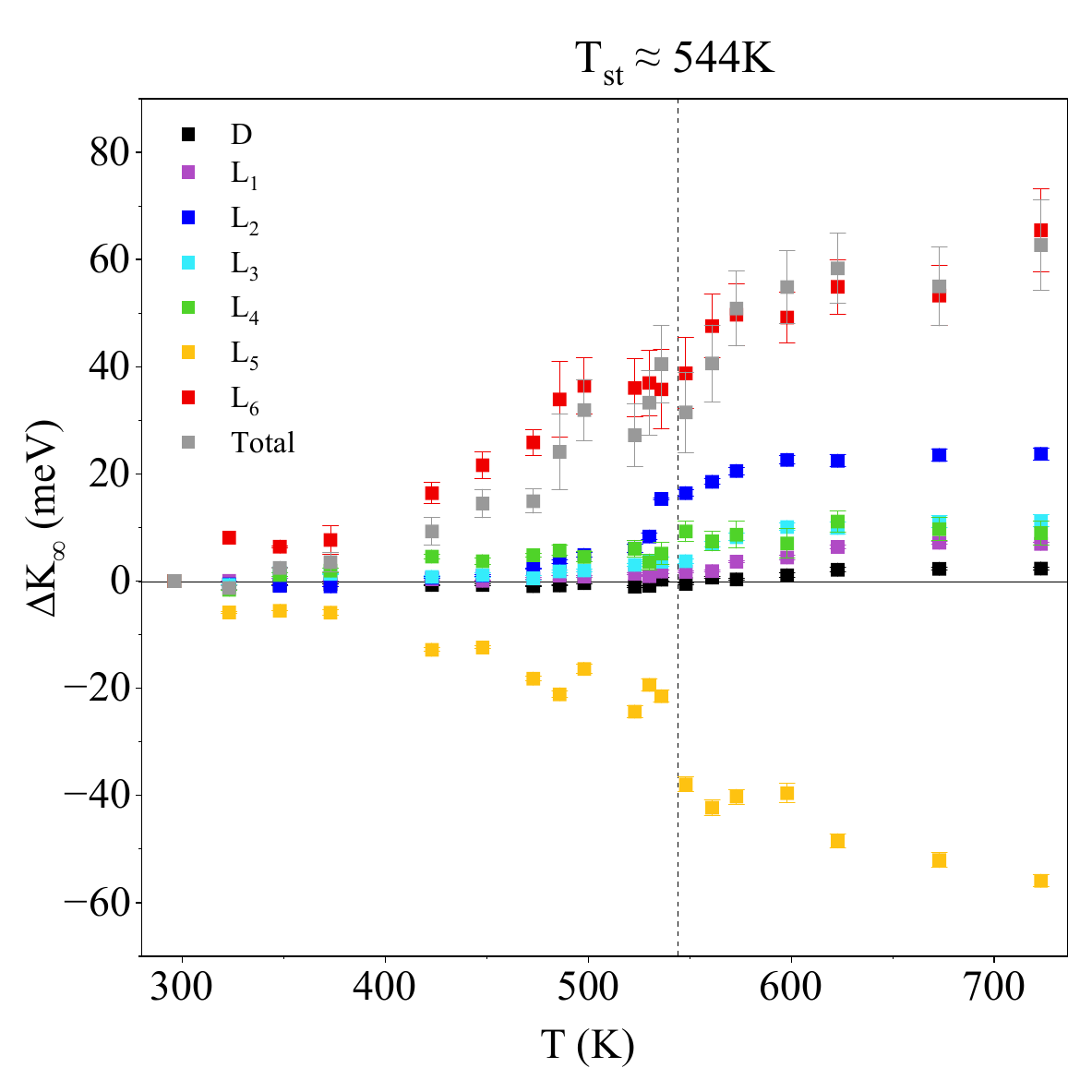}
\put(2,90){\textbf{d)}}
\end{overpic}
\end{minipage}

\caption{\label{fig:figS6} Component-resolved spectral-weight analysis of the fitted optical responses. (a) Integrated spectral weight $K_{\mathrm{int},j}(T)$ of the Drude term and the fitted Lorentz components, evaluated over the measured spectral range up to 26\,000~cm$^{-1}$ (see equation (\ref{eq4})). (b) Corresponding total spectral weight $K_{\infty,j}(T)$ enclosed inside each component (see equation (\ref{eq5})). (c) Spectral-weight variation within the measured range relative to 296~K, $\Delta K_{\mathrm{int},j}(T)=K_{\mathrm{int},j}(T)-K_{\mathrm{int},j}(296~\mathrm{K})$. (d) Corresponding variation of the total component spectral weight, $\Delta K_{\infty,j}(T)=K_{\infty,j}(T)-K_{\infty,j}(296~\mathrm{K})$. The vertical dashed lines mark the structural transition at $T_{\mathrm{st}}\simeq544$~K.}
\end{figure*}